\pdfoutput=1

\documentclass[11pt,twoside,a4paper,cmspaper,final,collab]{cms-tdr}
\def\svnVersion{a7d73eb-D}\def\svnDate{2019/10/01}\def\cmsCernNoTag{CERN-EP-2019-018}\def\cmsCernDate{\today}\def\cmsMessage{"Published in the European Physical Journal C as \href{http://dx.doi.org/10.1140/epjc/s10052-019-7387-y}{\doi{10.1140/epjc/s10052-019-7387-y}.}"}

\begin{document}\cmsNoteHeader{TOP-17-020}

\hyphenation{had-ron-i-za-tion}
\hyphenation{cal-or-i-me-ter}
\hyphenation{de-vices}
\RCS$HeadURL$
\RCS$Id$
\newlength\cmsFigWidth
\newlength\cmsTabSkip\setlength{\cmsTabSkip}{1ex}
\ifthenelse{\boolean{cms@external}}{\setlength\cmsFigWidth{0.85\columnwidth}}{\setlength\cmsFigWidth{0.4\textwidth}}
\ifthenelse{\boolean{cms@external}}{\providecommand{\cmsLeft}{top\xspace}}{\providecommand{\cmsLeft}{left\xspace}}
\ifthenelse{\boolean{cms@external}}{\providecommand{\cmsRight}{bottom\xspace}}{\providecommand{\cmsRight}{right\xspace}}
\cmsNoteHeader{TOP-17-020}
\ifthenelse{\boolean{cms@external}}{\providecommand{\cmsTable}[1]{#1}}{\providecommand{\cmsTable}[1]{\resizebox{\textwidth}{!}{#1}}}
\title{Search for new physics in top quark production in dilepton final states in proton-proton collisions at $\sqrt{s}$ = 13\TeV}
\titlerunning{Search for new physics via top quark production in dilepton final states at 13\TeV}

\date{\today}
\abstract{
A search for new physics in top quark production is performed in proton-proton collisions at 13\TeV.
The data set corresponds to an integrated luminosity of 35.9\fbinv collected in 2016 with the CMS detector.
Events with two opposite-sign isolated leptons (electrons or muons), and {\cPqb} quark jets in the final state are selected.
The search is sensitive to new physics in top quark pair production and in single top quark production in association with a {\PW} boson.
No significant deviation from the standard model expectation is observed. Results are interpreted in the framework of effective field theory and constraints on the relevant effective couplings are set, one at a time, using a dedicated multivariate analysis. 
This analysis differs from previous searches for new physics in the top quark sector by explicitly separating $\cPqt\PW$ from \ttbar events and exploiting the specific sensitivity of the $\cPqt\PW$ process to new physics.
}
\hypersetup{
pdfauthor={CMS Collaboration},
pdftitle={Search for new physics in top quark production in dilepton final states in proton-proton collisions at sqrt(s) = 13 TeV},
pdfsubject={CMS},
pdfkeywords={CMS, physics, top, dileptons}}
\maketitle

\section{Introduction}
\label{Int}

Because of its large mass, close to the electroweak (EW) symmetry breaking scale, the top quark is predicted to play an important role in several new physics scenarios.
If the new physics scale is in the available energy range of the CERN LHC, the existence of new physics could be directly observed via the production of new particles.
Otherwise, new physics could affect standard model (SM) interactions indirectly, through modifications of SM couplings or enhancements of rare SM processes.
In this case, it is useful to introduce a model independent approach to parametrize and constrain possible deviations from SM predictions, independently of the fundamental theory of new physics.

Several searches for new physics in the top quark sector including new non-SM couplings of the top quark have been performed at the Tevatron and LHC colliders~\cite{Abazov:2012uga,Aaboud:2017aqp,Aaboud:2016hsq,Khachatryan:2016sib,Khachatryan:2014vma,CMS:2014bea,Abazov:2010qk,Aaltonen:2008qr,Khachatryan:2016sib,Aad:2015gea,Khachatryan:2016kzg}.
Most of the previous analyses followed the anomalous coupling approach in which the SM Lagrangian is extended for possible new interactions.
Another powerful framework to parametrize deviations with respect to the SM prediction is the effective field theory (EFT) \cite{Buchmuller:1985jz,Grzadkowski:2010es}.
Constraints obtained on anomalous couplings can be translated to the effective coupling bounds~\cite{Zhang:2010dr,Abazov:2012uga}.
Several groups have performed global fits of top quark EFT to unfolded experimental data from the Tevatron and LHC colliders~\cite{Buckley:2015lku,Hartland:2019bjb}.
Due to the limited access to data and details of the associated  uncertainties, correlations between various cross section measurements and related uncertainties are neglected in a global fit on various unfolded measurements. 
On the other hand, EFT operators could affect backgrounds for some processes constructively or destructively while cross sections are measured with the SM assumptions for background processes.
Inside the CMS Collaboration and with direct access to data, all mentioned points could be considered properly.
 
\begin{figure*}[t!b]
\centering\includegraphics[width=1\textwidth]{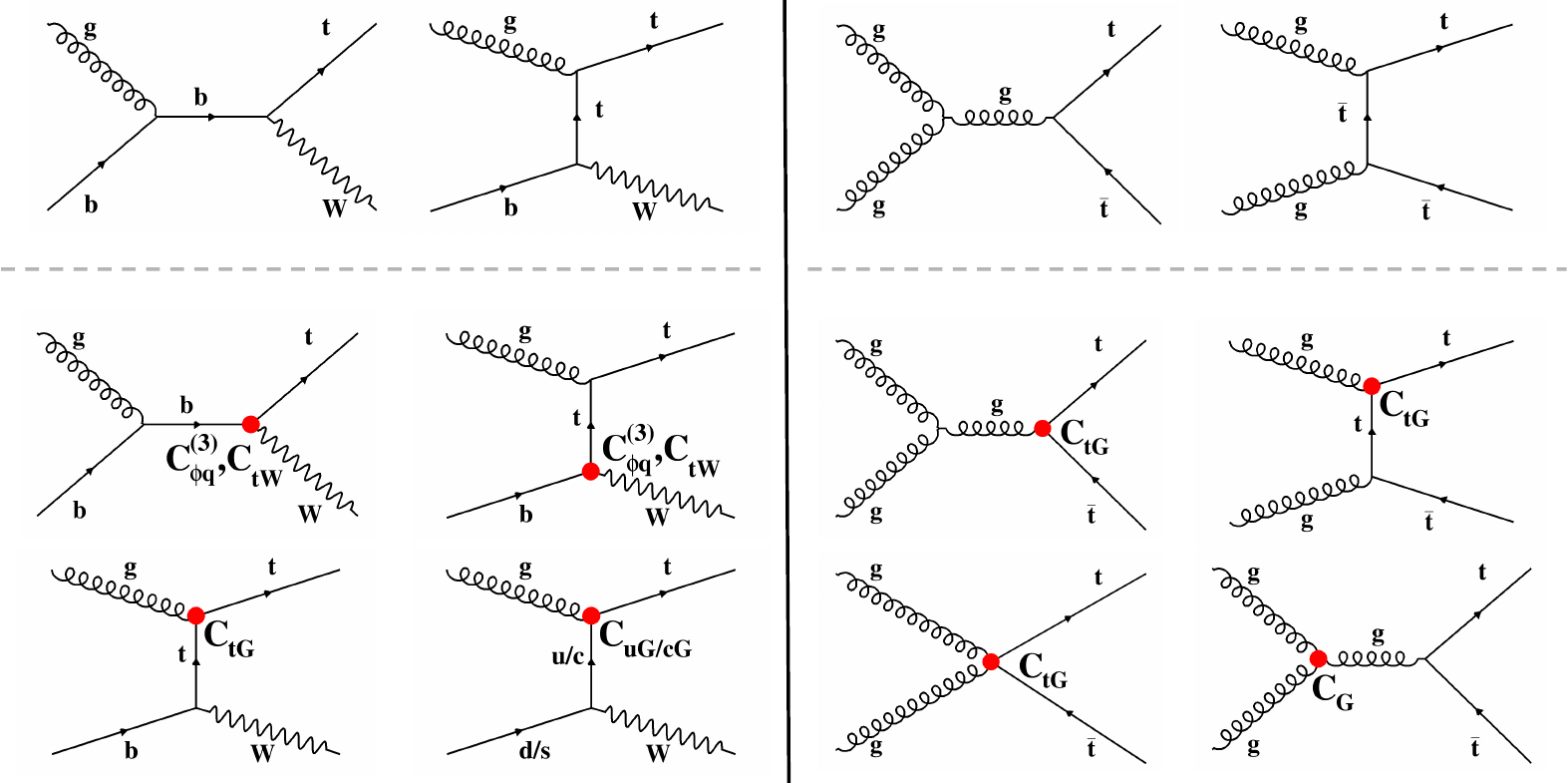}
\caption{Representative Feynman diagrams for the $\cPqt\PW$ (left panel) and \ttbar (right panel) production at leading order. The upper row presents the SM diagrams, the middle and lower rows  present diagrams corresponding to the $O_{\phi \cPq}^{(3)}$, $O_{\cPqt\PW}$,$O_{\cPqt\cPG}$, $O_{\cPG}$ and $O_{\mathrm {u/cG}}$ contributions.}\label{fig-feyn}
\end{figure*}

In this paper,  the EFT approach is followed to search for new physics in the top quark sector in the dilepton final states.
In Refs.~\cite{Zhang:2010dr,Durieux:2014xla}, all dimension-six operators that contribute to top quark pair (\ttbar) production and single top quark production in association with a {\PW} boson ($\cPqt\PW$) are investigated.
The operators and the related effective Lagrangians, which are relevant for dilepton final states, can be written as~\cite{Grzadkowski:2010es}:
\begin{linenomath}
\ifthenelse{\boolean{cms@external}}
{
\begin{multline}
\label{eq1}
O_{\phi \cPq}^{(3)} = (\phi^+\Pgt^iD_\Pgm\phi)(\Paq\cPgg^\Pgm\Pgt^i\cPq), \\
L_{\mathrm {eff}}=\frac{{\mathrm {C}}_{\phi \cPq}^{(3)}}{{\sqrt 2}\Lambda^2} \Pg v^2\cPaqb\cPgg^\Pgm P_{\mathrm {L}}\cPqt\PW^-_\Pgm+\text{h.c.},
\end{multline}
\begin{multline}
\label{eq2}
O_{\cPqt\PW} = (\Paq\sigma^{\Pgm\PGn}\Pgt^i\cPqt)\tilde{\phi}\PW^i_{\Pgm\PGn},  \\
L_{\mathrm {eff}}=-2\frac{{\mathrm {C}}_{\cPqt\PW}}{\Lambda^2}v\cPaqb\sigma^{\Pgm\PGn}P_{\mathrm {R}}\cPqt \partial_\cPgn \PW^-_\Pgm+\text{h.c.},
\end{multline}
\begin{multline}
\label{eq3}
O_{\cPqt\cPG} = (\Paq\sigma^{\Pgm\PGn}\Lam^a \cPqt)\tilde{\phi}\cPG^a_{\Pgm\PGn}, \\
L_{\mathrm {eff}}= \frac{{\mathrm {C}}_{\cPqt\cPG}}{{\sqrt 2}\Lambda^2}v\left(\cPaqt\sigma^{\Pgm\PGn}\Lam^a \cPqt\right) \cPG_{\Pgm\PGn}^a+\text{h.c.},
\end{multline}
\begin{multline}
\label{eq4}
O_{\cPG} = f_{abc} \cPG^{a\cPgn}_{\Pgm} \cPG^{b\PGr}_{\cPgn} \cPG^{\cPqc \Pgm}_{\PGr},  \\
L_{\mathrm {eff}}= \frac{{\mathrm {C}}_{\cPG}}{\Lambda^2} f_{abc}\cPG^{a\cPgn}_\Pgm \cPG^{b\PGr}_\cPgn \cPG^{\cPqc\Pgm}_{\PGr},
\end{multline}
\begin{multline}
\label{eq5}
O_{\cPqu(\cPqc)\cPG} =  (\Paq\sigma^{\Pgm\PGn}\Lam^a \cPqt)\tilde{\phi}\cPG^a_{\Pgm\PGn}, \\
L_{\mathrm {eff}}= \frac{{\mathrm {C}}_{\cPqu(\cPqc)\cPG}}{{\sqrt 2}\Lambda^2}v\left(\cPaqu \left(\cPaqc\right)\sigma^{\Pgm\PGn}\Lam^a \cPqt\right) \cPG_{\Pgm\PGn}^a+\text{h.c.},
\end{multline}
}
{
\begin{align}
&O_{\phi \cPq}^{(3)} = (\phi^+\Pgt^iD_\Pgm\phi)(\Paq\cPgg^\Pgm\Pgt^i\cPq)  ,&& L_{\mathrm {eff}}=\frac{{\mathrm {C}}_{\phi \cPq}^{(3)}}{{\sqrt 2}\Lambda^2} \Pg v^2\cPaqb\cPgg^\Pgm P_{\mathrm {L}}\cPqt\PW^-_\Pgm+\text{h.c.}, \label{eq1}\\
&O_{\cPqt\PW} = (\Paq\sigma^{\Pgm\PGn}\Pgt^i\cPqt)\tilde{\phi}\PW^i_{\Pgm\PGn} ,&& L_{\mathrm {eff}}=-2\frac{{\mathrm {C}}_{\cPqt\PW}}{\Lambda^2}v\cPaqb\sigma^{\Pgm\PGn}P_{\mathrm {R}}\cPqt \partial_\cPgn \PW^-_\Pgm+\text{h.c.}, \label{eq2}\\
&O_{\cPqt\cPG} = (\Paq\sigma^{\Pgm\PGn}\Lam^a \cPqt)\tilde{\phi}\cPG^a_{\Pgm\PGn} ,&& L_{\mathrm {eff}}= \frac{{\mathrm {C}}_{\cPqt\cPG}}{{\sqrt 2}\Lambda^2}v\left(\cPaqt\sigma^{\Pgm\PGn}\Lam^a \cPqt\right) \cPG_{\Pgm\PGn}^a+\text{h.c.}, \label{eq3} \\
&O_{\cPG} = f_{abc}\cPG^{a\cPgn}_\Pgm \cPG^{b\PGr}_\cPgn \cPG^{\cPqc\Pgm}_\PGr ,&& L_{\mathrm {eff}}= \frac{{\mathrm {C}}_{\cPG}}{\Lambda^2} f_{abc}\cPG^{a\cPgn}_\Pgm \cPG^{b\PGr}_\cPgn \cPG^{\cPqc\Pgm}_\PGr, \label{eq4}\\
&O_{\cPqu(\cPqc)\cPG} =  (\Paq\sigma^{\Pgm\PGn}\Lam^a \cPqt)\tilde{\phi}\cPG^a_{\Pgm\PGn}  ,&& L_{\mathrm {eff}}= \frac{{\mathrm {C}}_{\cPqu(\cPqc)\cPG}}{{\sqrt 2}\Lambda^2}v\left(\cPaqu \left(\cPaqc\right)\sigma^{\Pgm\PGn}\Lam^a \cPqt\right) \cPG_{\Pgm\PGn}^a+\text{h.c.}, \label{eq5}
\end{align}
}
\end{linenomath}
where $D_\Pgm=\partial_\Pgm-i\Pg_s\frac{1}{2}\Lam^a\cPG_\Pgm^a-i\Pg\frac{1}{2}\Pgt^i\PW_\Pgm^i-i\Pg^{'}YB_\Pgm$, $\PW_{\Pgm\PGn}^i=\partial_\Pgm \PW_\cPgn^i-\partial_\cPgn \PW^i_\Pgm+\Pg\epsilon_{ijk}\PW^j_\Pgm \PW^k_\cPgn$, $\cPG_{\Pgm\PGn}^a=\partial_\Pgm \cPG_\cPgn^a-\partial_\cPgn \cPG^a_\Pgm+\Pg_sf^{abc}\cPG^b_\Pgm \cPG^c_\cPgn$, $\sigma^{\Pgm\PGn} = \frac{1}{2}[\cPgg^{\Pgm},\cPgg^{\cPgn}]$, $P_{\mathrm{L,R}}=\frac{1}{2} (1\mp \cPgg^5)$, and the symbols $\cPq$, $\cPqt$ and $\phi$ ($\tilde{\phi}=\epsilon\phi^*$) in the operators represent  the left-handed quark doublet, the right-handed top quark singlet, and the Higgs boson doublet fields, respectively.
The parameters C$_{\phi \cPq}^{(3)}$, C$_{\cPqt\PW}$, C$_{\cPqt\cPG}$, C$_{\cPG}$ and C$_{\cPqu(\cPqc)\cPG}$ stand for the dimensionless Wilson coefficients, also called effective couplings.
The variable $\Lambda$ represents the energy scale beyond which new physics becomes relevant. A detailed description of the operators is given in~Refs. \cite{Zhang:2010dr,Durieux:2014xla,AguilarSaavedra:2018nen}.
In this analysis, four-fermion operators involved in \ttbar~production are not probed. 
Up to order $\Lambda^{-2}$, the $\cPqt\PW$ and \ttbar production cross sections and most of the differential observables considered in this analysis do not receive CP-odd contributions. Therefore, we only probe CP-even operators with real coefficients.
The operators $O_{\phi \cPq}^{(3)}$ and $O_{\cPqt\PW}$ modify the SM interaction between the {\PW} boson, top quark, and {\cPqb} quark ($\PW\cPqt\cPqb$).
We consider the EFT effects in the production of top quarks not in their decays \cite{Godbole:2018wfy}.
The operator $O_{\cPqt\cPG}$ is called the chromomagnetic dipole moment operator of the top quark and can arise from various models of new physics \cite{Stange:1993td,Li:1995fj}.
The triple-gluon field strength operator $O_{\cPG}$ represents the only genuinely gluonic CP conserving term that can appear at dimension six within an effective strong interaction Lagrangian.
Although it is shown that jet production at the LHC can set a tight constraint on the C$_{\cPG}$ \cite{Krauss:2016ely}, \ttbar production is also considered as a promising channel~\cite{Cho:1994yu,Hirschi:2018etq}.
The operators $O_{\cPqu\cPG}$ and $O_{\cPqc\cPG}$ lead to flavor-changing neutral current (FCNC) interactions of the top quark and contribute to $\cPqt\PW$ production.
The effect of introducing new couplings C$_{\phi \cPq}^{(3)}$, C$_{\cPqt\PW}$, C$_{\cPqt\cPG}$ and C$_{\cPqu(\cPqc)\cPG}$ can be investigated in $\cPqt\PW$ production.
The chromomagnetic dipole moment operator of the top quark also affects \ttbar production. In the case of C$_{\cPG}$ coupling, only \ttbar production is modified.
It should be noted that the $O_{\cPqt\PW}$ and $O_{\cPqt\cPG}$ operators with imaginary coefficients lead to CP-violating effects.
Representative Feynman diagrams for SM and new physics contributions in $\cPqt\PW$ and \ttbar production are shown in Fig.~\ref{fig-feyn}.

A variety of limits have been set on the $\PW\cPqt\cPqb$ anomalous coupling through single top quark $t$-channel production
and measurements of the {\PW} boson polarization from top quark decay by the D0~\cite{Abazov:2012uga},
ATLAS \cite{Aaboud:2017aqp,Aaboud:2016hsq} and CMS \cite{Khachatryan:2016sib,Khachatryan:2014vma} Collaborations.
Direct limits on the top quark chromomagnetic dipole moment have been obtained  by the CMS Collaboration at 7  and 13\TeV using top quark pair production events~\cite{CMS:2014bea,Khachatryan:2016kzg}.
Searches for top quark FCNC interactions have been performed at the Tevatron~\cite{Abazov:2010qk,Aaltonen:2008qr} and LHC~\cite{Khachatryan:2016sib,Aad:2015gea}
via single top quark production and limits are set on related anomalous couplings.

In this paper, a search for new physics in top quark production using an EFT framework is reported.
This is the first such search for new physics that uses the $\cPqt\PW$ process.
Final states with two opposite-sign isolated leptons (electrons or muons) in association with jets identified as originating from the fragmentation of a bottom quark (``\cPqb jets'') are analyzed.
The search is sensitive to new physics contributions to $\cPqt\PW$ and \ttbar production, and the six effective couplings, C$_{\cPG}$, C$_{\phi \cPq}^{(3)}$, C$_{\cPqt\PW}$, C$_{\cPqt\cPG}$, C$_{\cPqu\cPG}$, and C$_{\cPqc\cPG}$, are constrained assuming one non-zero effective coupling at a time.
The effective couplings affect both the rate of \ttbar and $\cPqt\PW$ production and the kinematic distributions of final state particles.
For the C$_{\phi \cPq}^{(3)}$, C$_{\cPqt\PW}$, C$_{\cPqt\cPG}$, and C$_{\cPG}$ effective couplings, the deviation from the SM prediction is dominated by the interference term between SM and new physics diagrams, which is linear with respect to the effective coupling. Therefore, the kinematic distributions of the final-state particles vary as a function of the Wilson coefficients. For  small  effective couplings the kinematic distributions approach those predicted by the SM.
On the other hand, the new physics terms due to the C$_{\cPqu\cPG}$ and C$_{\cPqc\cPG}$ effective couplings do not interfere with the SM $\cPqt\PW$ process, and the kinematic distributions of final-state particles are determined by the new physics terms independently of the SM prediction.
In this analysis, we use the rates of $\cPqt\PW$ and \ttbar production to probe the C$_{\phi \cPq}^{(3)}$, C$_{\cPqt\PW}$, C$_{\cPqt\cPG}$, and C$_{\cPG}$ effective couplings. Variations in both rate and kinematic distributions of final-state particles are employed to probe the C$_{\cPqu\cPG}$ and C$_{\cPqc\cPG}$ effective couplings.
The analysis utilizes proton-proton ($\Pp\Pp$) collision data collected by the CMS experiment in 2016 at a center-of-mass energy of 13\TeV, corresponding to an integrated luminosity of 35.9\fbinv.

The paper is structured as follows. In Section~\ref{cms}, a description of the CMS detector is given and
the simulated samples used in the analysis are detailed.
The event selection and the SM background estimation are presented in Section~\ref{object}.
Section~\ref{signal} presents a description of the signal extraction procedure.
An overview of the systematic uncertainty treatment is given in Section~\ref{sys}.
Finally, the constraints on the effective couplings are presented in Section~\ref{result}, and a summary is given in Section~\ref{summary}.

\section{The CMS detector and event simulation}
\label{cms}

The central feature of the CMS apparatus is a superconducting solenoid of 6\unit{m} internal diameter, providing a magnetic field of 3.8\unit{T}. Within the solenoid volume are a silicon pixel and strip tracker, a lead tungstate crystal electromagnetic calorimeter (ECAL), and a brass and scintillator hadron calorimeter, each composed of a barrel and two endcap sections. Forward calorimeters extend the pseudorapidity ($\eta$) coverage provided by the barrel and endcap detectors. Muons are detected in gas-ionisation chambers embedded in the steel flux-return yoke outside the solenoid.
A more detailed description of the CMS detector, together with a definition of the coordinate system used and the relevant kinematic variables, can be found in Ref.~\cite{Chatrchyan:2008zzk}.

The Monte Carlo (MC) samples for the \ttbar, $\cPqt\PW$ and diboson ($\mathrm{VV} = \PW\PW$, $\PW\PZ$, $\PZ\PZ$) SM processes are simulated
using the \textsc{Powheg-Box} event generator (v1 for $\cPqt\PW$, v2 for \ttbar and diboson)~\cite{POWHEG,POWHEG2,POWHEG:tW,POWHEG:SingleTop} at the next-to-leading order (NLO),
interfaced with \PYTHIA (v8.205)~\cite{Sjostrand:2014zea} to simulate parton showering
and to match soft radiations with the contributions from the matrix elements.
The \PYTHIA tune CUETP8M1 \cite{Khachatryan:2015pea} is used for all samples except for the \ttbar sample, for which the tune CUETP8M2~\cite{CMS-PAS-TOP-16-021} is used.
The NNPDF3.0~\cite{Ball:2014uwa} set of the parton distribution functions (PDFs) is used.
The \ttbar and $\cPqt\PW$ samples are normalized to the next-to-next-to-leading order (NNLO) and approximate NNLO  cross section calculations, respectively~\cite{Czakon:2011xx,Kidonakis:2015nna}.
In order to better describe the transverse momentum (\pt) distribution of the top quark in \ttbar events, the top quark \pt spectrum simulated with \POWHEG  is reweighted to match the differential top quark \pt distribution at NNLO Quantum ChromoDynamics (QCD) accuracy and including EW corrections calculated in Ref. \cite{Czakon:2017wor}.
Other SM background contributions, from Drell--Yan (DY), \ttbar+V, \ttbar+$\cPgg$, and $\PW+\cPgg$ processes, are simulated at NLO using the \MGvATNLO (v2.2.2) event generator~\cite{MADGRAPH,Alwall:2014hca,Frederix:2012ps}, interfaced with \PYTHIA v8 for parton showering and hadronization.
The events include the effects of additional $\Pp\Pp$ interactions in the same or nearby bunch crossings (pileup) and are
weighted according to the observed pileup distribution in the analyzed data.
The CMS detector response is simulated using \GEANTfour~(v9.4) \cite{GEANT4_1,GEANT4_2}, followed by a detailed trigger simulation.  Simulated events are reconstructed with the same algorithms as used for data.

In order to calculate the total cross sections for the \ttbar and $\cPqt\PW$ processes and generate events in the presence of new effective interactions, the operators of Eqs.~\ref{eq1}--\ref{eq5} have been implemented in the universal \textsc{FeynRules} output (UFO) format \cite{Degrande:2011ua} through the
\textsc{FeynRules} package \cite{Alloul:2013bka}. The output EFT model is used in the \MGvATNLO (v2.2.2) event generator~\cite{MADGRAPH,Alwall:2014hca}.
If we allow for the presence of one operator at a time, the total cross section up to $\mathcal{O}(\Lambda^{-4})$ can be parametrized as
\begin{equation}
\label{eq2x}
        \sigma=\sigma_{\mathrm {SM}}+
        {\mathrm {C}}_{i}\sigma_i^{(1)}+
    {\mathrm {C}}_{i}^2\sigma_{i}^{(2)},
\end{equation}
where the C$_i$s are effective  couplings introduced in Eqs.~\ref{eq1}--\ref{eq5}. Here, $\sigma_i^{(1)}$ is the contribution to the cross section due to the interference term between the SM diagrams and diagrams with one EFT vertex.   The cross section $\sigma_i^{(2)}$ is the pure new physics  contribution.
We use the most precise available SM cross section prediction, which are $\sigma_{\text{SM}}^{\ttbar}=832^{+20}_{-29}\,\text{(scales)}\pm 35\,(\text{PDF}+\alpS)\unit{pb}$ and $\sigma_{\text{SM}}^{\cPqt\PW}=71.7\pm 1.8\,\text{(scales)}\pm 3.4\,(\text{PDF}+\alpS)\unit{pb}$ for \ttbar and $\cPqt\PW$ production, respectively \cite{Czakon:2011xx,Kidonakis:2015nna}, where the $\alpS$ is strong coupling constant.
The first uncertainty reflects the uncertainties  in  the  factorization and  renormalization scales.
In the framework of EFT, the $\sigma_i^{(1)}$  and $\sigma_i^{(2)}$ terms have been calculated at NLO accuracy for all of the operators, except $O_{\cPG}$~\cite{Franzosi:2015osa,Zhang:2016omx,Durieux:2014xla}. 
At the time the work for this paper was concluded, there was no available UFO model including the $O_{\cPG}$ operator at the NLO.
The values of $\sigma_i^{(1)}$ and $\sigma_i^{(2)}$ for various effective couplings at LO and available $K$ factors are given in Table \ref{xsVScoupling}.

\begin{table*}[htb]
\topcaption{Contribution to the cross section due to the interference between the SM diagrams and diagrams with one EFT vertex ($\sigma_i^{(1)}$), and the pure new physics ($\sigma_i^{(2)}$) for \ttbar and $\cPqt\PW$ production [in\unit{pb}] for the various effective couplings for $\Lambda$  = 1\TeV. The respective $K$ factors ($\sigma_i^{\mathrm {NLO}}/\sigma_i^{\mathrm {LO}}$) are also shown.
}
\label{xsVScoupling}
\centering
\begin{tabular}{llllllll}
\hline
Channel & Contribution & $C_{\cPG}$ & $C_{\phi \cPq}^{(3)}$ & $C_{\cPqt\PW}$ & $C_{\cPqt\cPG}$ & $C_{\cPqu\cPG}$ & $C_{\cPqc\cPG}$   \\
\hline
\multirow{4}{*}{\ttbar} &   $\sigma_i^{(1) \mathrm { -LO}}$  &  31.9\unit{pb}   & \multicolumn{1}{c}{\NA}   & \multicolumn{1}{c}{\NA}   & 137\unit{pb}  & \multicolumn{1}{c}{\NA} & \multicolumn{1}{c}{\NA}\\
 &   $K^{(1)}$  &  \multicolumn{1}{c}{\NA}  & \multicolumn{1}{c}{\NA}   & \multicolumn{1}{c}{\NA}   & 1.48  & \multicolumn{1}{c}{\NA}   & \multicolumn{1}{c}{\NA}             \\
 &   $\sigma_i^{(2)\mathrm{ -LO}}$  &  102.3\unit{pb}  & \multicolumn{1}{c}{\NA}   & \multicolumn{1}{c}{\NA}   & 16.4\unit{pb}   & \multicolumn{1}{c}{\NA}   & \multicolumn{1}{c}{\NA}             \\
 &   $K^{(2)}$  &  \multicolumn{1}{c}{\NA}   & \multicolumn{1}{c}{\NA}  & \multicolumn{1}{c}{\NA}   & 1.44   & \multicolumn{1}{c}{\NA}   & \multicolumn{1}{c}{\NA}             \\[\cmsTabSkip]
\multirow{4}{*}{\cPqt\PW} &   $\sigma_i^{(1) \mathrm { -LO}}$  &  \multicolumn{1}{c}{\NA}   & 6.7\unit{pb}  & $-$4.5\unit{pb}   & 3.3\unit{pb}  & 0  & 0             \\
 &   $K^{(1)}$ &  \multicolumn{1}{c}{\NA}   & 1.32     & 1.27   & 1.27    & 0   & 0             \\
 &   $\sigma_i^{(2) \mathrm { -LO}}$  &  \multicolumn{1}{c}{\NA}   & 0.2\unit{pb}  & 1\unit{pb}   & 1.2\unit{pb}  & 16.2\unit{pb}  & 4.6\unit{pb}            \\
 &   $K^{(2)}$ &  \multicolumn{1}{c}{\NA}   & 1.31    & 1.18   & 1.06   & 1.27   &1.27  \\
 \hline
\end{tabular}
\end{table*}

\section{Event selection and background estimation}
\label{object}

The event selection for this analysis is similar to the one used in Ref.~\cite{Khachatryan:2016kzg}.
The events of interest are recorded by the CMS detector using a combination of dilepton and single-lepton triggers.
Single-lepton triggers require at least one isolated electron (muon) with $\pt>27\,(24)\GeV$.
The dilepton triggers select events with at least two leptons with loose isolation requirements and \pt for the leading
and sub-leading leptons greater than 23 and 12 (17 and 8)\GeV for the $\Pe\Pe$ ($\Pgm\Pgm$) final state.
In the $\Pe\Pgm$ final state, in the case of the leading lepton being an electron, the events are triggered if the electron-muon pair has a \pt greater than 23 and 8\GeV for the electron and muon, respectively. In the case of the leading lepton being a muon, the trigger thresholds are 23 and 12\GeV for the muon and electron, respectively~\cite{Khachatryan:2016bia}.

Offline, the particle-flow (PF) algorithm \cite{Sirunyan:2017ulk} aims to reconstruct and identify each individual particle
with an optimized combination of information from the various elements of the CMS detector.
Electron candidates are reconstructed using tracking and ECAL information~\cite{Khachatryan:2015hwa}. Requirements on electron identification variables based on shower shape
and track-cluster matching are further applied to the reconstructed electron candidates, together with isolation criteria~\cite{Khachatryan:2016kzg,Khachatryan:2015hwa}.
Electron candidates are selected with $\pt>20\GeV$ and $\abs{\eta}< 2.4$.
Electron candidates within the range $1.444 < \abs{\eta}< 1.566$, which corresponds to the transition region between the barrel
and endcap regions of the ECAL, are not considered.
Information from the tracker and the muon spectrometer are combined in a global fit to reconstruct muon candidates~\cite{Chatrchyan:2012xi}.
Muon candidates are further required to have a high-quality fit including a minimum number of hits in both systems, and to be isolated~\cite{Khachatryan:2016kzg,Chatrchyan:2012xi}.
The muons used in this analysis are selected inside the fiducial region of the muon spectrometer, $\abs{\eta}<2.4$, with a minimum \pt of 20\GeV.

The PF candidates are clustered into jets using the anti-\kt algorithm with a distance parameter of 0.4~\cite{Cacciari:2011ma,Cacciari:2008gp,Cacciari:2005hq}.
Jets are calibrated in data and simulation, accounting for energy deposits of particles from pileup \cite{Khachatryan:2016kdb}.
Jets with $\pt>30\GeV$ and $\abs{\eta} <2.4$ are selected;
loose jets are defined as jets with the \pt range between 20 and 30\GeV.
Jets originating from the hadronization of {\cPqb} quarks are identified using the combined secondary vertex algorithm \cite{Sirunyan:2017ezt};
this algorithm combines information from track impact parameters and secondary vertices identified within a given jet.
The chosen working point provides a signal identification efficiency of approximately 68\%  with a probability  to misidentify light-flavor jets as \cPqb jets of approximately 1\%~\cite{Sirunyan:2017ezt}.
The missing transverse momentum vector \ptvecmiss
is defined as the projection on the plane perpendicular to the proton beams axis of the negative vector sum
of the momenta of all reconstructed PF candidates in the event~\cite{Khachatryan:2014gga}.
Corrections to the jet energies are propagated to \ptvecmiss.
Its magnitude is referred to as \ptmiss.

Events are required to have at least two leptons (electrons or muons) with opposite sign and
with an invariant mass above 20\GeV. The leading lepton must fulfill $\pt>25\GeV$.
For the same-flavor lepton channels, to suppress the DY background, the dilepton invariant mass must not be within 15\GeV of the {\PZ} boson mass and a minimal value (of 60\GeV) on \ptmiss is applied.

\begin{figure*}[!htb]
  \centering
\includegraphics[width=0.45\textwidth]{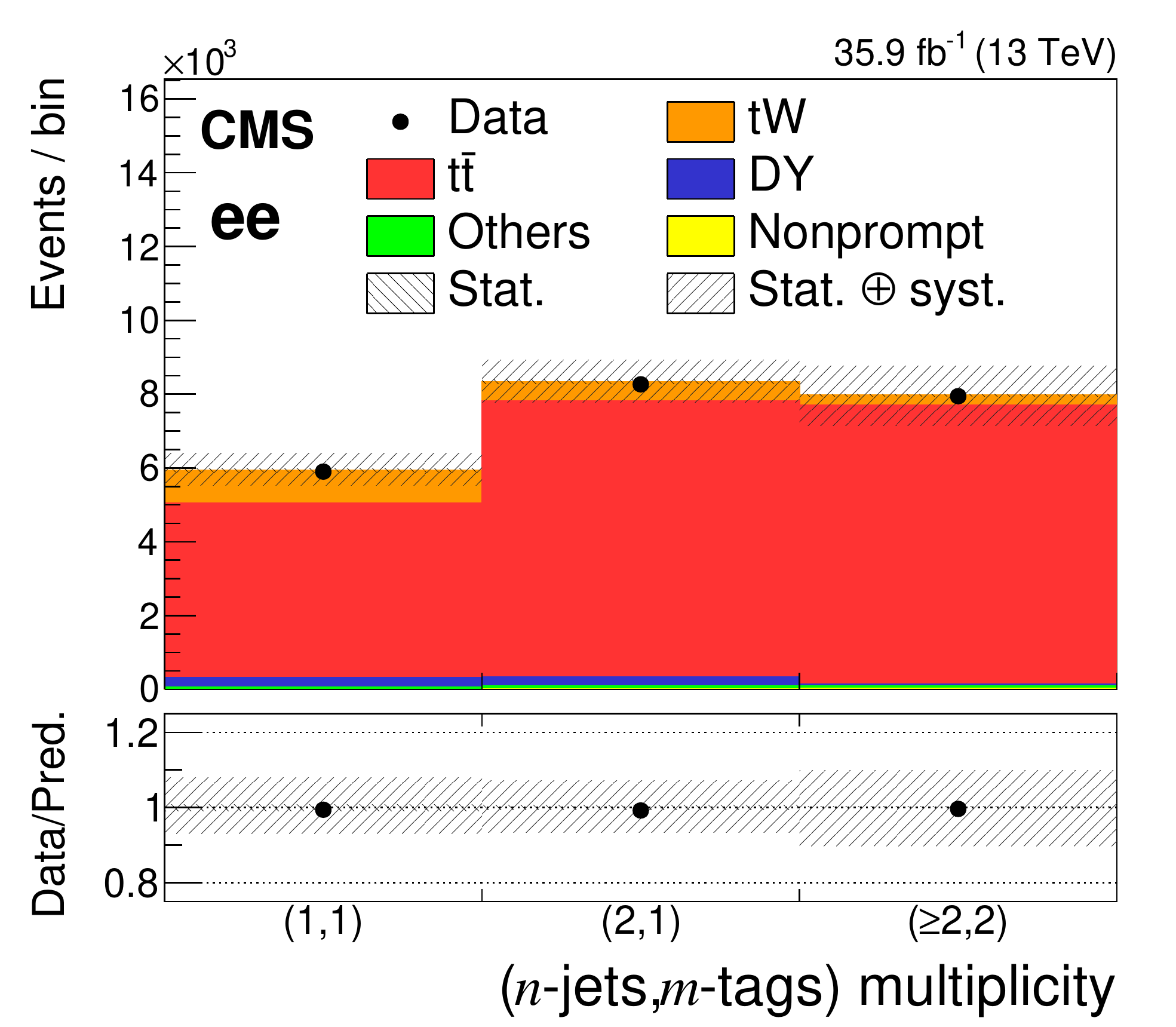}
\includegraphics[width=0.45\textwidth]{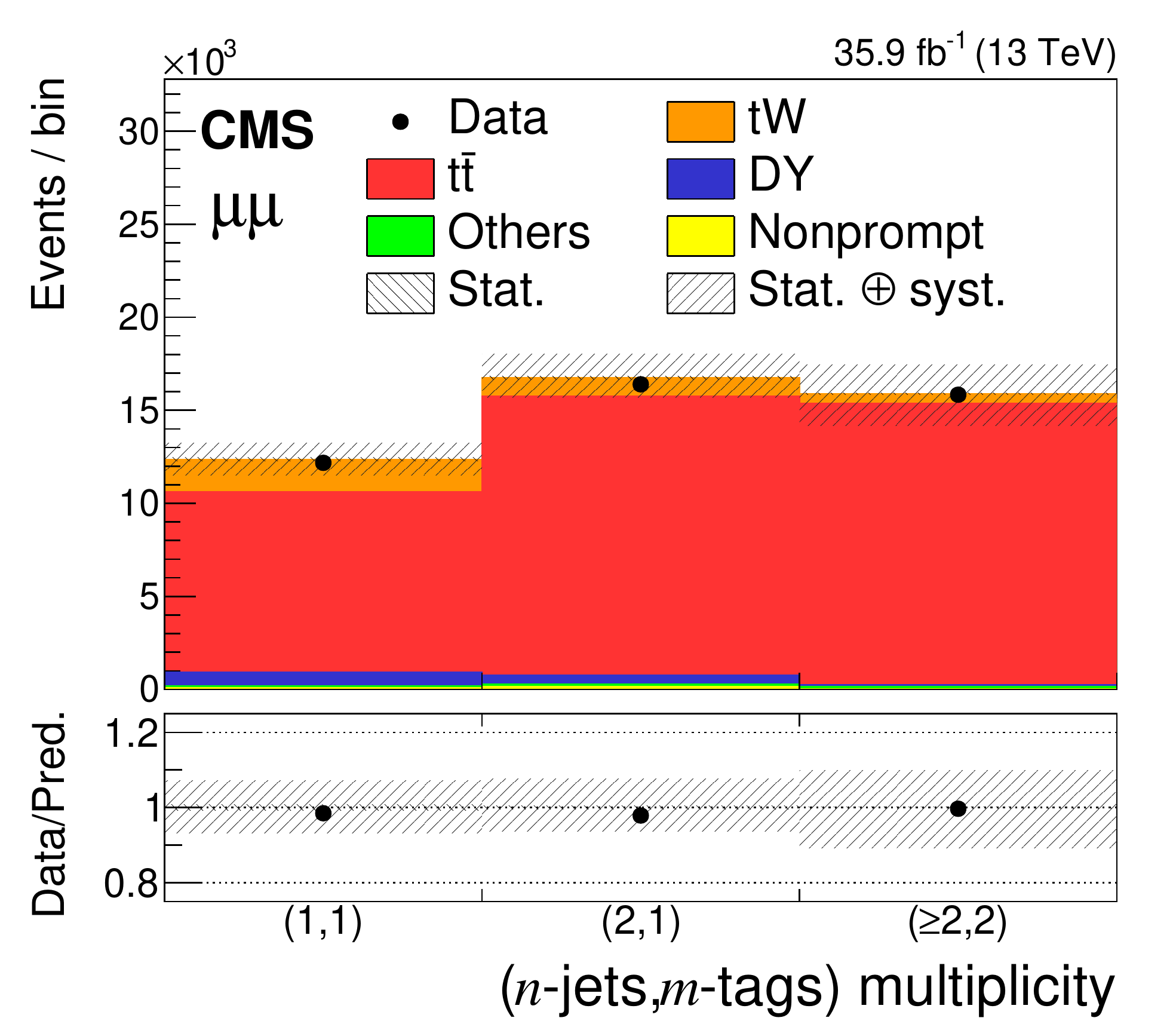}
\includegraphics[width=0.45\textwidth]{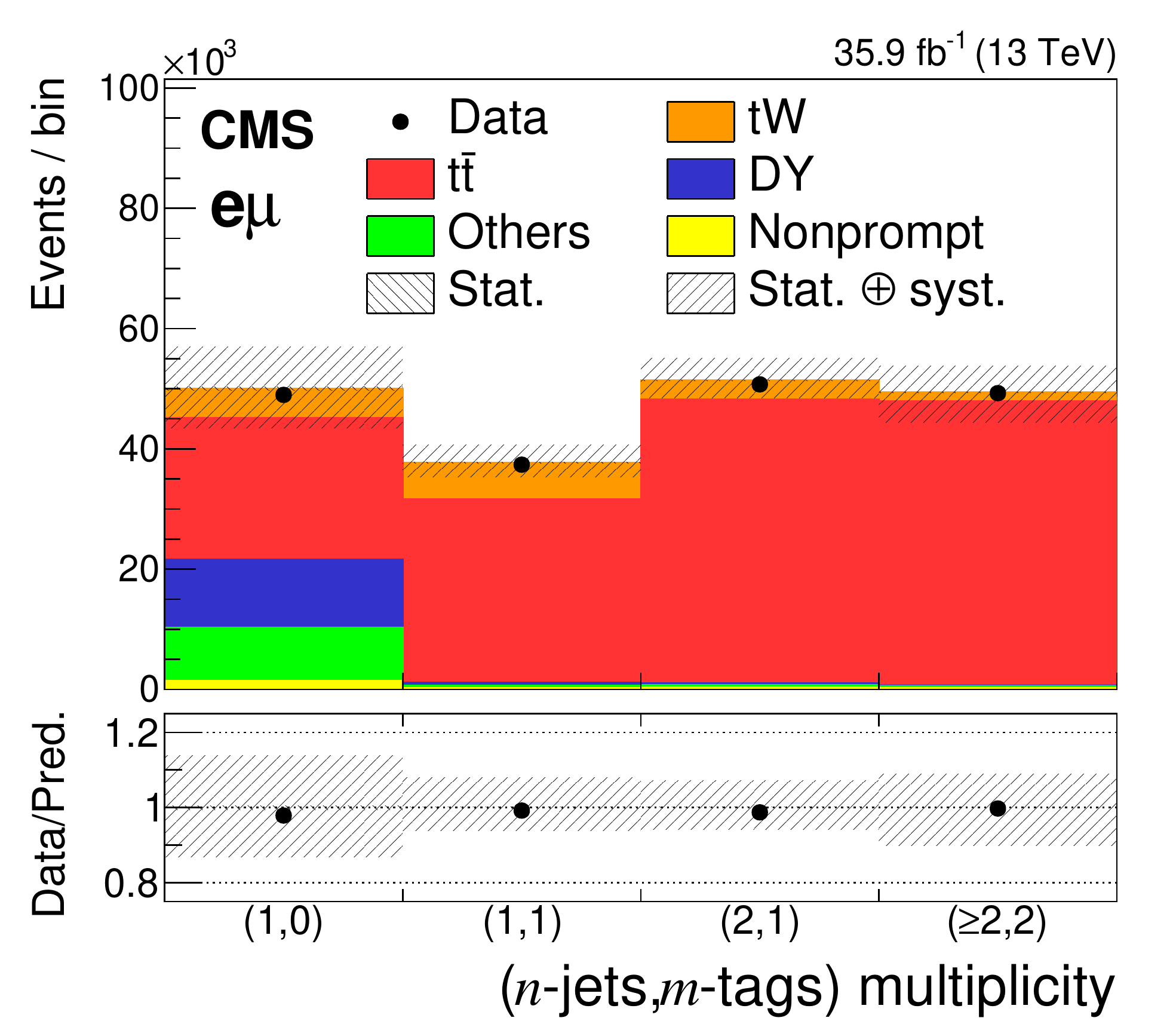}
    \caption{The observed number of events and SM background predictions in the search regions of the analysis for the $\Pe\Pe$ (upper left), $\Pgm\Pgm$  (upper right), and $\Pe\Pgm$ (lower) channels. The hatched bands correspond to the quadratic sum of statistical and systematic uncertainties in the event yield for the SM background predictions. The ratios of data to the sum of the predicted yields are shown at the bottom of each plot. The narrow hatched bands represent the contribution from the statistical uncertainty in the MC simulation.
    \label{fig:binee}}
\end{figure*}

The events are divided into the $\Pe\Pe$, $\Pe\Pgm$, and $\Pgm\Pgm$ channels according to the flavors of the two leptons with the highest \pt
and are further categorized in different bins depending on the number of jets (``$n$-jets'') and number {\cPqb}-tagged jets (``$m$-tags'') in the final state.
The largest number of  $\cPqt\PW$ events is expected in the category with exactly one  {\cPqb}-tagged jet (1-jet,1-tag) followed by the category with two jets, of which one a  \cPqb jet (2-jets,1-tag). Events in the categories with more than  two jets and exactly two {\cPqb}-tagged jets  are dominated by the \ttbar process ($\geq$2-jets,2-tags).
Categories with zero \cPqb jets are dominated by DY events in the $\Pe\Pe$ and $\Pgm\Pgm$ channels and are not used in the analysis. However, in the $\Pe\Pgm$ channel, the contamination of DY events is lower and a significant number of $\cPqt\PW$ events is present  in the  category with one jet and zero {\cPqb}-tagged jets (1-jet,0-tag). The latter category is included in this analysis.
In Fig. \ref{fig:binee}, the data in the ten search regions are shown together with the SM background predictions.

The contributions of SM processes leading to two prompt leptons in the final state are estimated from simulated samples and are normalised to the integrated luminosity of the data. These contributions originate mainly  from  \ttbar, $\cPqt\PW$ and DY production.
Other SM processes, such as diboson,  \ttbar+V and  \ttbar+$\cPgg$ have significantly smaller contributions.

To correct the DY simulation for the efficiency of the \ptmiss threshold and for the mismodeling of the heavy-flavour content, scale factors are derived using the ratio of the numbers of simulated events inside and outside the dilepton invariant mass window, 76 to 106\GeV.
The observed event yield inside the window is scaled to estimate the DY background outside the mass window~\cite{Khachatryan:2010ez}.

The nonprompt lepton backgrounds which contain fake lepton(s) from a misreconstructed $\cPgg$ or jet(s) are also considered.
The contribution of misidentified or converted $\cPgg$ events from the  $\PW\cPgg$ process is estimated from MC simulation.
The contribution from  $\PW$+jets and multijet processes is estimated by a data-based technique using events with same-sign leptons.
The method is based on the assumption that the probability of assigning positive or negative charge to the fake lepton is equal.
Therefore, the background contribution from fake leptons in the final selection (opposite-sign sample) can be estimated from the corresponding sample with same-sign leptons.
In this latter same-sign event sample, the remaining small contribution from prompt-lepton backgrounds is subtracted from data using MC samples.

After all selections, the expected numbers of events from $\cPqt\PW$, \ttbar, DY, and remaining background contributions mentioned above, as well as the total number of background events are reported in Table~\ref{tab:limit} for the $\Pe\Pe$, $\Pe\Pgm$, and $\Pgm\Pgm$ channels and for the various ($n$-jets,$m$-tags) categories.
We find generally very good agreement between data and predictions, within the uncertainties of the data.

\begin{table*}[!htb]
\centering
\topcaption{Number of expected events from $\cPqt\PW$, \ttbar and DY production, from the remaining  backgrounds (other), total background contribution and observed events in data after all selections for the $\Pe\Pe$, $\Pe\Pgm$, and $\Pgm\Pgm$ channels and for different ($n$-jets,$m$-tags) categories.
The uncertainties correspond to the statistical contribution only for the individual background predictions and to the quadratic sum of the statistical and systematic contributions for the total background predictions.
}
\cmsTable{
\begin{tabular}{ccr@{$\pm$}lr@{$\pm$}lr@{$\pm$}lr@{$\pm$}lr@{$\pm$}lc}
\hline
\multirow{2}{*}{Channel}  & \multirow{2}{*}{($n$-jets,$m$-tags)}  & \multicolumn{10}{c}{Prediction}   & \multirow{2}{*}{Data}     \\
                          &                                        & \multicolumn{2}{c}{\cPqt\PW}      & \multicolumn{2}{c}{\ttbar}     & \multicolumn{2}{c}{DY}  & \multicolumn{2}{c}{Other}        &  \multicolumn{2}{c}{Total yield}   &        \\ \hline
\multirow{3}{*}{$\Pe\Pe$}       & (1,1)     & 884\, &\,8    &  4741\,&\,15    &    258\,&\,50~       &   53\,&\,5       & 5936\,&\,470     & 5902      \\
                          & (2,1)     & 518\, &\,6    &  7479\,&\,19    &    241\,&\,53~       &   94\,&\,5       & 8331\,&\,597     & 8266      \\
                          & ($\ge$2,2)& 267\, &\,4    &  7561\,&\,18    &     46\,&\,24~       &   99\,&\,4       & 7973\,&\,819     & 7945      \\ [\cmsTabSkip]
\multirow{4}{*}{$\Pe\Pgm$}   & (1,0)     & 4835\,&\,20  &  23557\,&\,35    &  11352\,&\,277       &10294\,&\,72     & 50038\,&\,6931   & 48973     \\
                          & (1,1)     & 6048\,&\,22  &  30436\,&\,38    &    561\,&\,66~       &  629\,&\,13     & 37673\,&\,2984   & 37370     \\
                          & (2,1)     & 3117\,&\,16  &  47206\,&\,48    &    278\,&\,48~       &  781\,&\,9      & 51382\,&\,3714   & 50725     \\
                          & ($\ge$2,2)& 1450\,&\,10  &  47310\,&\,46    &     32\,&\,22~       &  598\,&\,9      & 49391\,&\,5010   & 49262     \\ [\cmsTabSkip]
\multirow{3}{*}{$\Pgm\Pgm$} & (1,1)     & 1738\,&\,12  &   9700\,&\,21    &    744\,&\,90~       &  183\,&\,5      & 12366\,&\,879    & 12178     \\
                          & (2,1)     & 989\, &\,9    & 14987\,&\,27    &    501\,&\,75~       &  275\,&\,5      & 16751\,&\,1276   & 16395     \\
                          & ($\ge$2,2)& 508\, &\,6    & 15136\,&\,26    &     82\,&\,24~       &  163\,&\,5      & 15889\,&\,1714   & 15838     \\ \hline
\end{tabular}}
\label{tab:limit}
\end{table*}

\section{Signal extraction using neural networks tools}
\label{signal}
The purpose of the analysis is to search for deviations from the SM predictions in the $\cPqt\PW$ and \ttbar production due to new physics, parametrized with the presence of new effective couplings. In order to investigate the effect of the non-zero effective couplings,
it is important to find suitable variables with high discrimination power between the signal and the  background.
Depending on the couplings, the total yield or the distribution of the output of a neural network (NN) algorithm is employed, as summarized in Table~\ref{tab:NNreg}.
The NN algorithm used in this analysis is a multilayer perceptron~\cite{Denby:1992jd}.

\begin{table*}[ht]
\centering
\topcaption{Summary of the observables used to probe the effective couplings  in various ($n$-jets,$m$-tags) categories in the $\Pe\Pe$, $\Pe\Pgm$, and $\Pgm\Pgm$ channels.}
\cmsTable{
\begin{tabular}{llccccc}
\hline
\multirow{2}{*}{Eff. coupling}         & \multirow{2}{*}{Channel}            &  \multicolumn{5}{c}{Categories } \\
& & 1-jet ,0-tag               & 1-jet ,1-tag             & 2-jets,1-tag & $>$2-jets ,1-tag & $\ge$2-jets,2-tags\\ \hline
\multirow{3}{*}{C$_{\cPG}$}      & $\Pe\Pe$         &  \NA  & Yield&Yield& \NA &Yield \\
                              & $\Pe\Pgm$     & Yield&Yield&Yield& \NA &Yield \\
                              & $\Pgm\Pgm$   &  \NA  &Yield&Yield& \NA &Yield \\[\cmsTabSkip]
\multirow{3}{*}{C$_{\phi \cPq}^{(3)}$, C$_{\cPqt\PW}$, C$_{\cPqt\cPG}$ } & $\Pe\Pe$         &  \NA  &NN$_{11}$ & NN$_{21}$ & \NA &Yield \\
                              & $\Pe\Pgm$     & NN$_{10}$ &NN$_{11}$ & NN$_{21}$& \NA &Yield \\
                              & $\Pgm\Pgm$   &  \NA  &NN$_{11}$ & NN$_{21}$& \NA &Yield \\[\cmsTabSkip]
\multirow{3}{*}{C$_{\cPqu\cPG}$, C$_{\cPqc\cPG}$} & $\Pe\Pe$         & \NA &\multicolumn{3}{c}{NN$_{\text{FCNC}}$}& \NA  \\
                              & $\Pe\Pgm$     & \NA &\multicolumn{3}{c}{NN$_{\text{FCNC}}$}& \NA  \\
                              & $\Pgm\Pgm$   & \NA &\multicolumn{3}{c}{NN$_{\text{FCNC}}$}& \NA \\
\hline
\end{tabular}}
\label{tab:NNreg}
\end{table*}

All the effective couplings introduced in Section \ref{Int} can contribute to $\cPqt\PW$ production except the triple gluon field strength operator, $O_{\cPG}$ which only affects \ttbar production.
As observed in previous analysis \cite{Cho:1994yu} and confirmed here, the top quark \pt distribution is sensitive to the triple-gluon field-strength operator.
The kinematic distributions of final-state particles show less discrimination power than the  top quark \pt distribution. In addition,  they vary with the value of C$_{\cPG}$ and approach the SM prediction for decreasing values of C$_{\cPG}$. Therefore, we use the total yield in  various categories to constrain the C$_{\cPG}$ effective coupling.

\begin{figure*}[!htbp]
  \centering
      \includegraphics[width=0.45\textwidth]{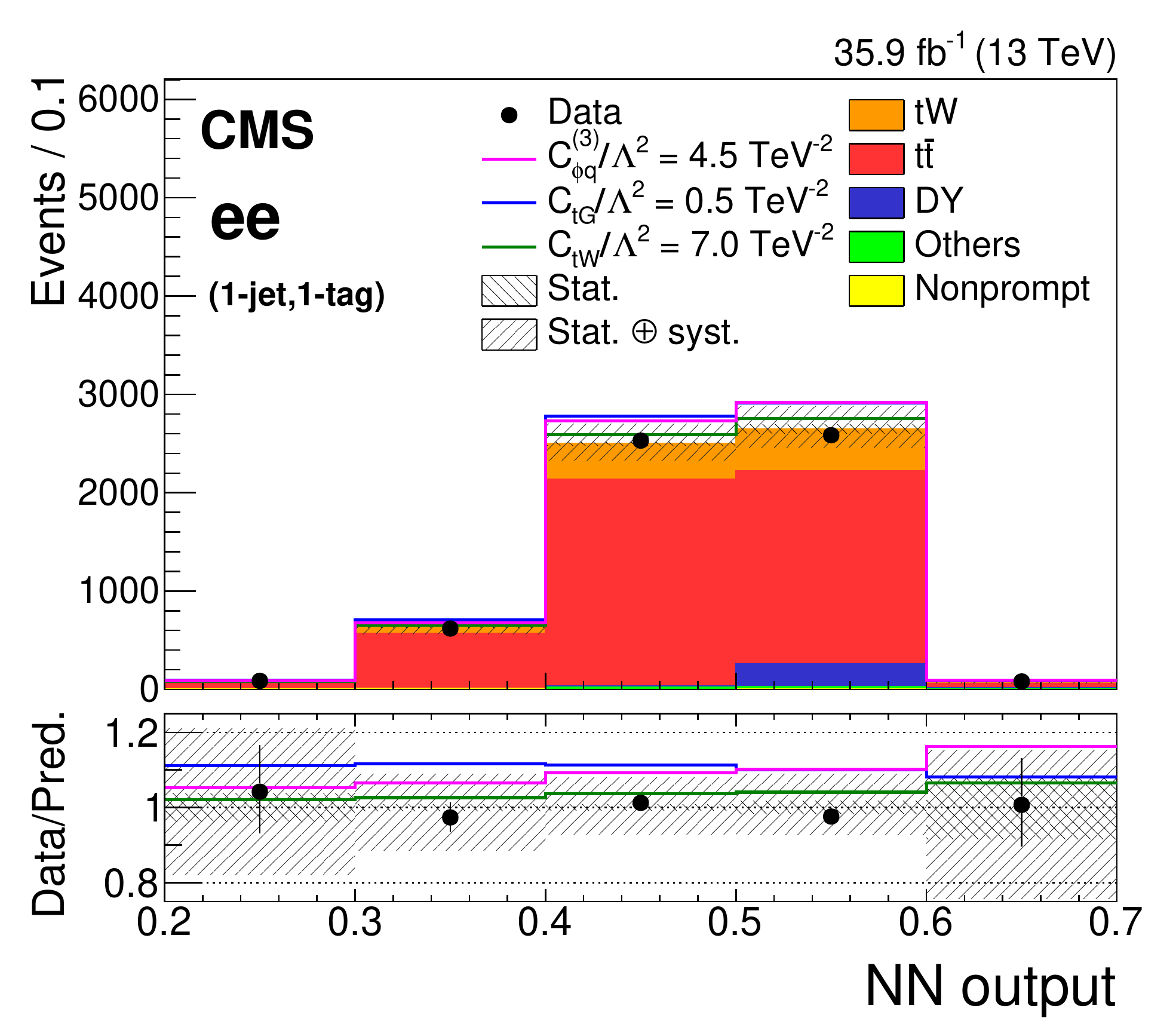} 
      \includegraphics[width=0.45\textwidth]{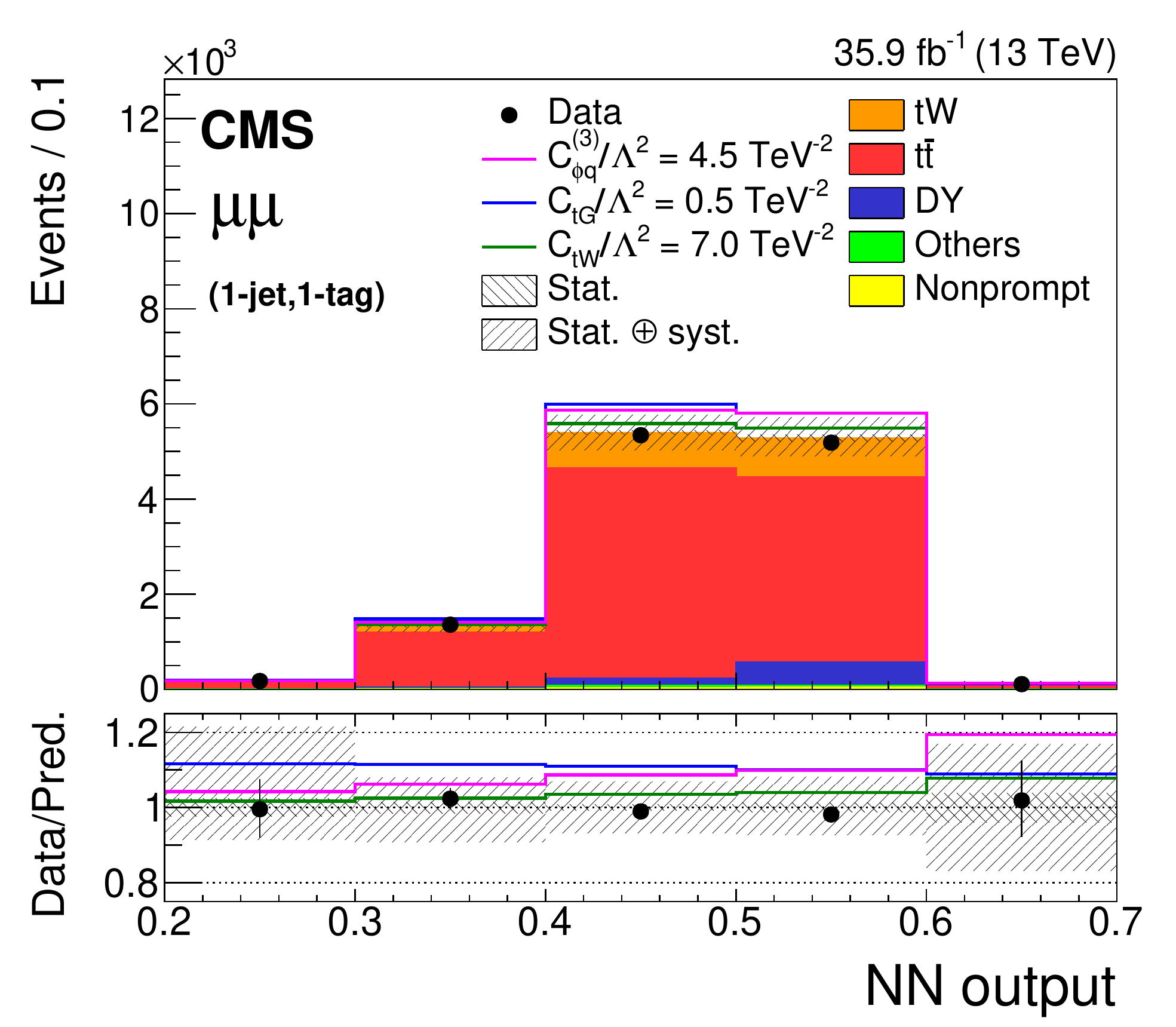}
      \includegraphics[width=0.45\textwidth]{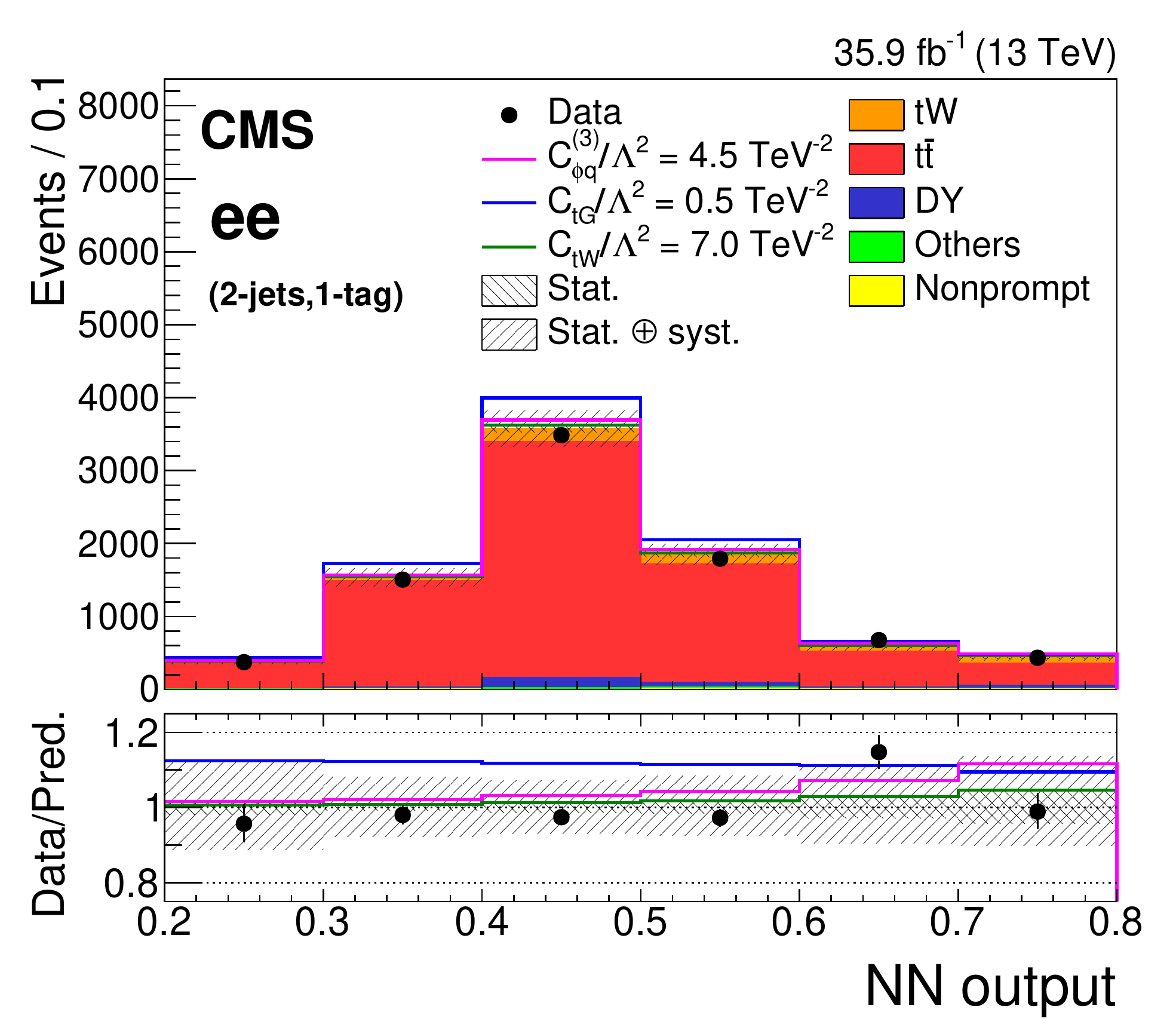} 
      \includegraphics[width=0.45\textwidth]{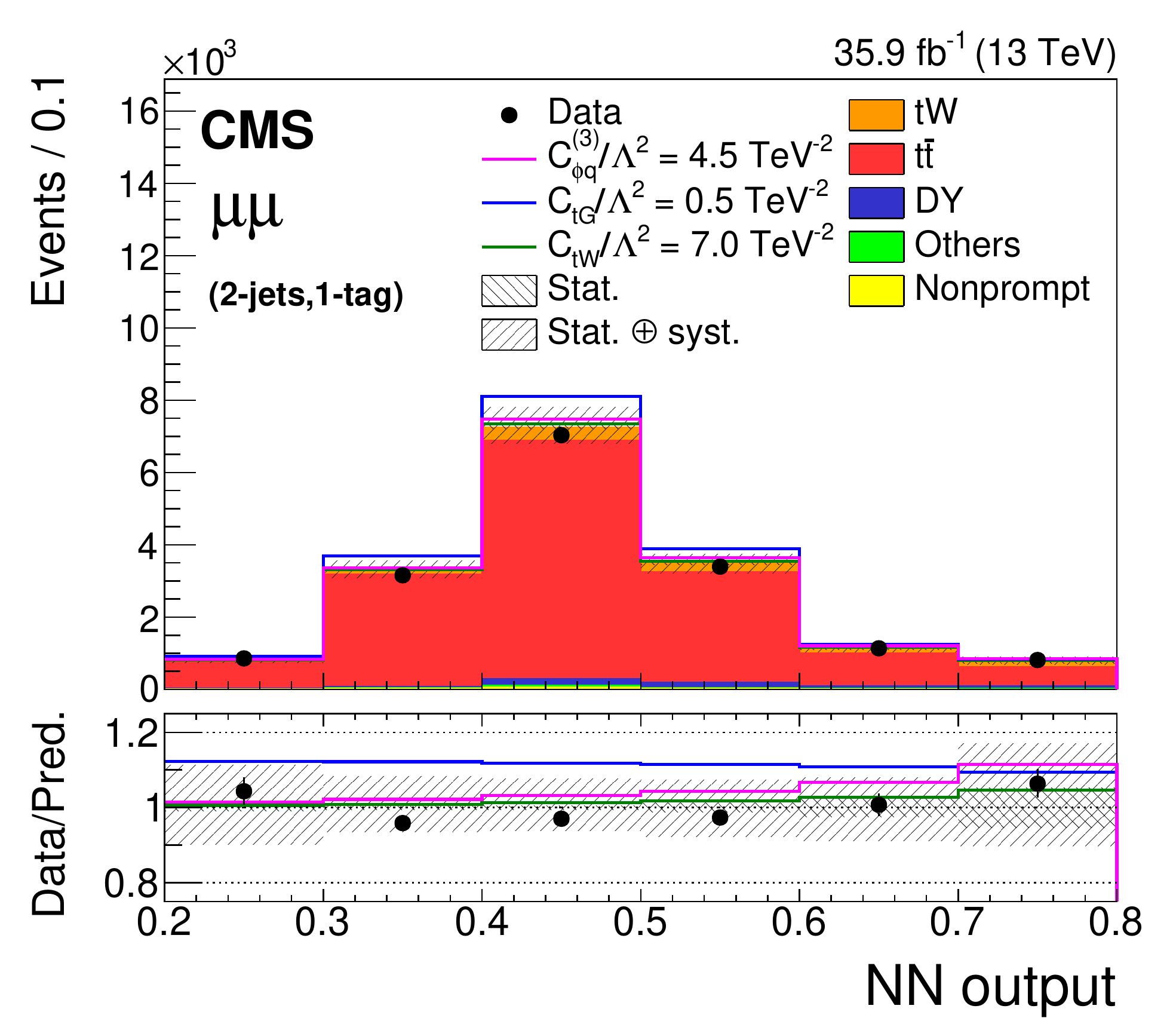}
    \caption{The NN output distributions for data and simulation for the $\Pe\Pe$ (left) and $\Pgm\Pgm$  (right) channels in 1-jet, 1-tag (upper) and 2-jets, 1-tag  (lower) categories. The hatched bands correspond to the quadratic sum of the statistical and systematic uncertainties in the event yield for the sum of signal and background predictions. The ratios of data to the sum of the predicted yields are shown at the lower panel of each graph. The narrow hatched bands represent the contribution from the statistical uncertainty in the MC simulation. In each plot, the expected distributions assuming specific values for the effective couplings (given in the legend) are shown as the solid curves.
    \label{fig:limit_binee1}}
\end{figure*}

\begin{figure*}[!htbp]
  \centering
      \includegraphics[width=0.45\textwidth]{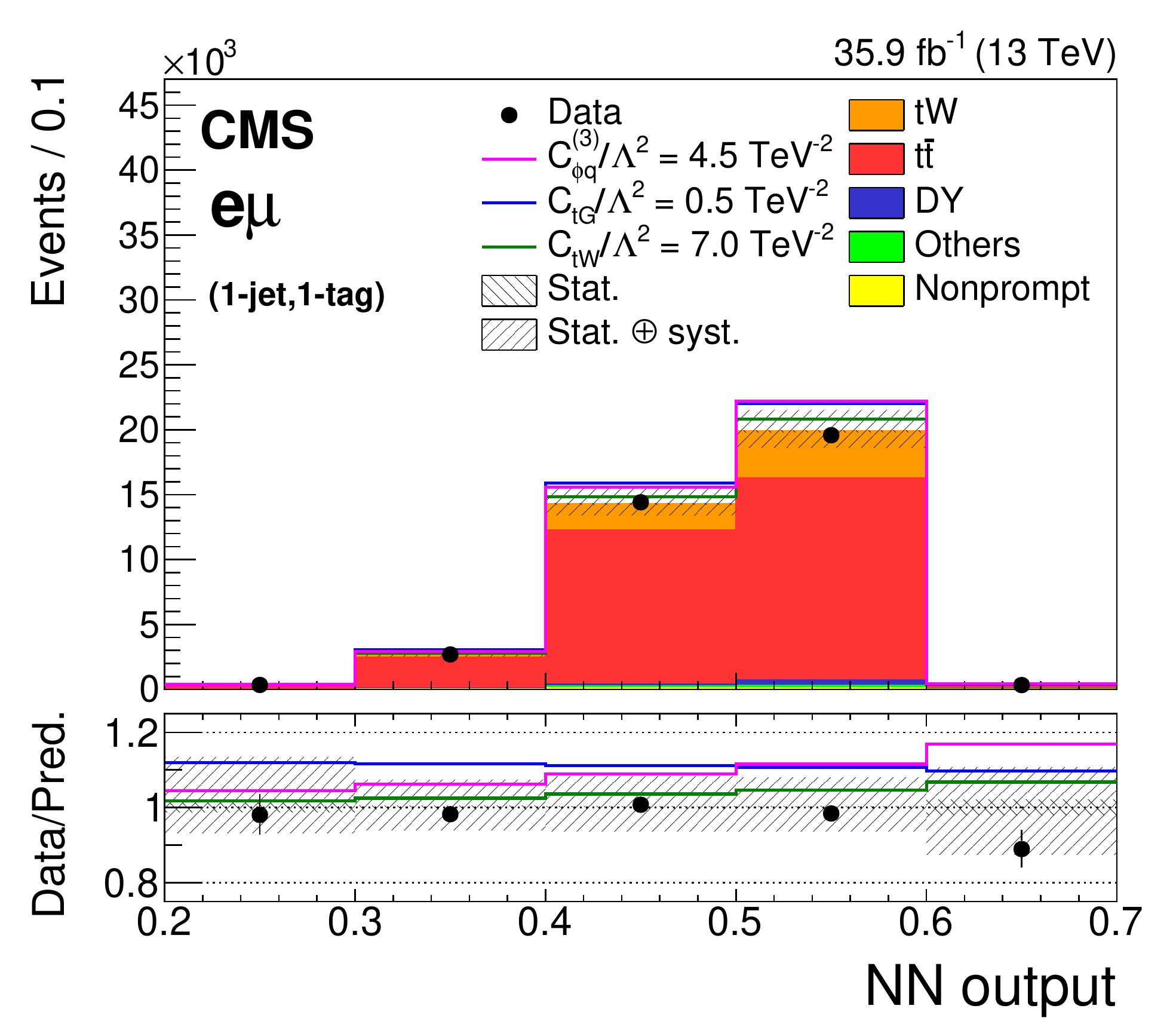}
      \includegraphics[width=0.45\textwidth]{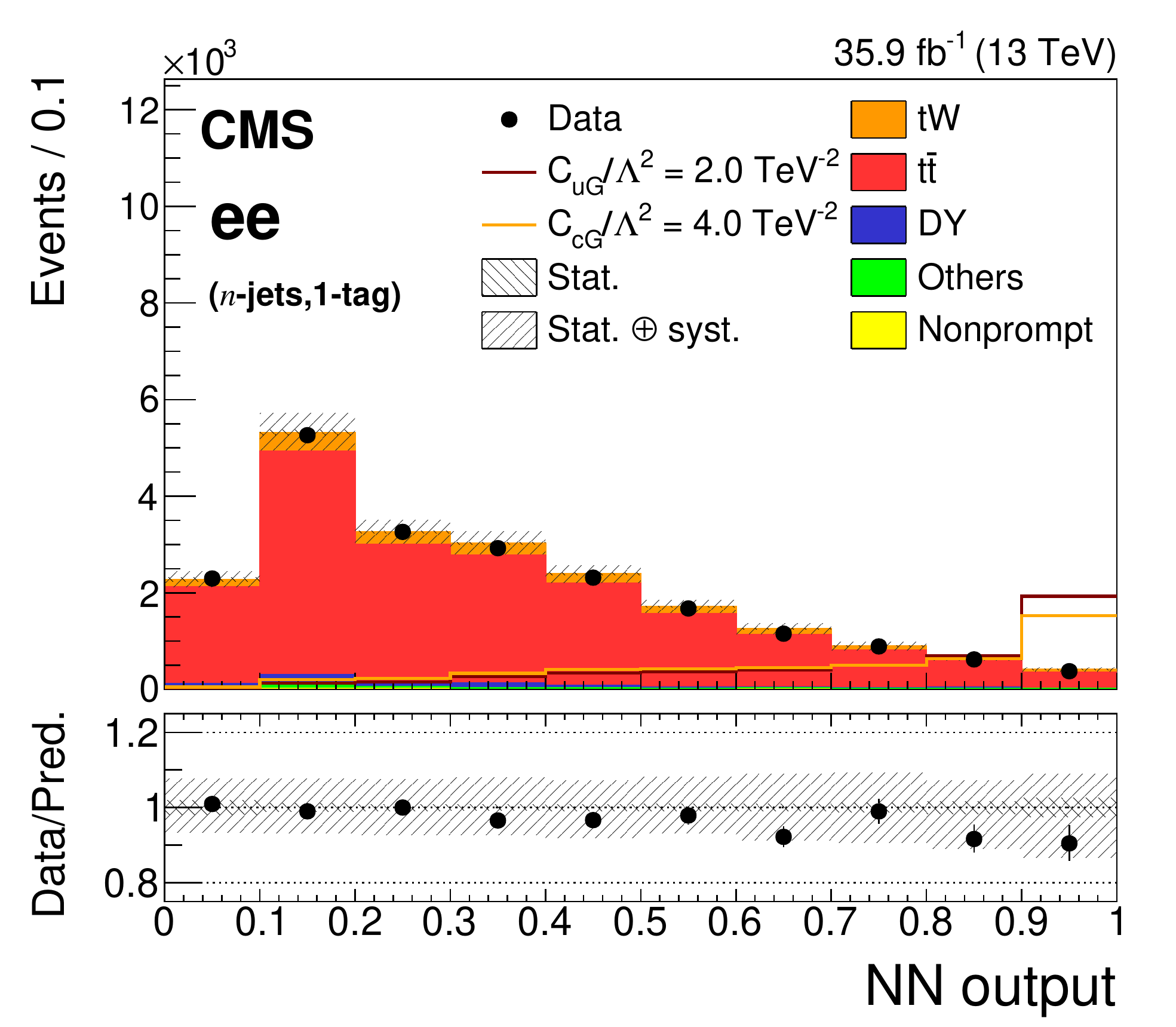}
      \includegraphics[width=0.45\textwidth]{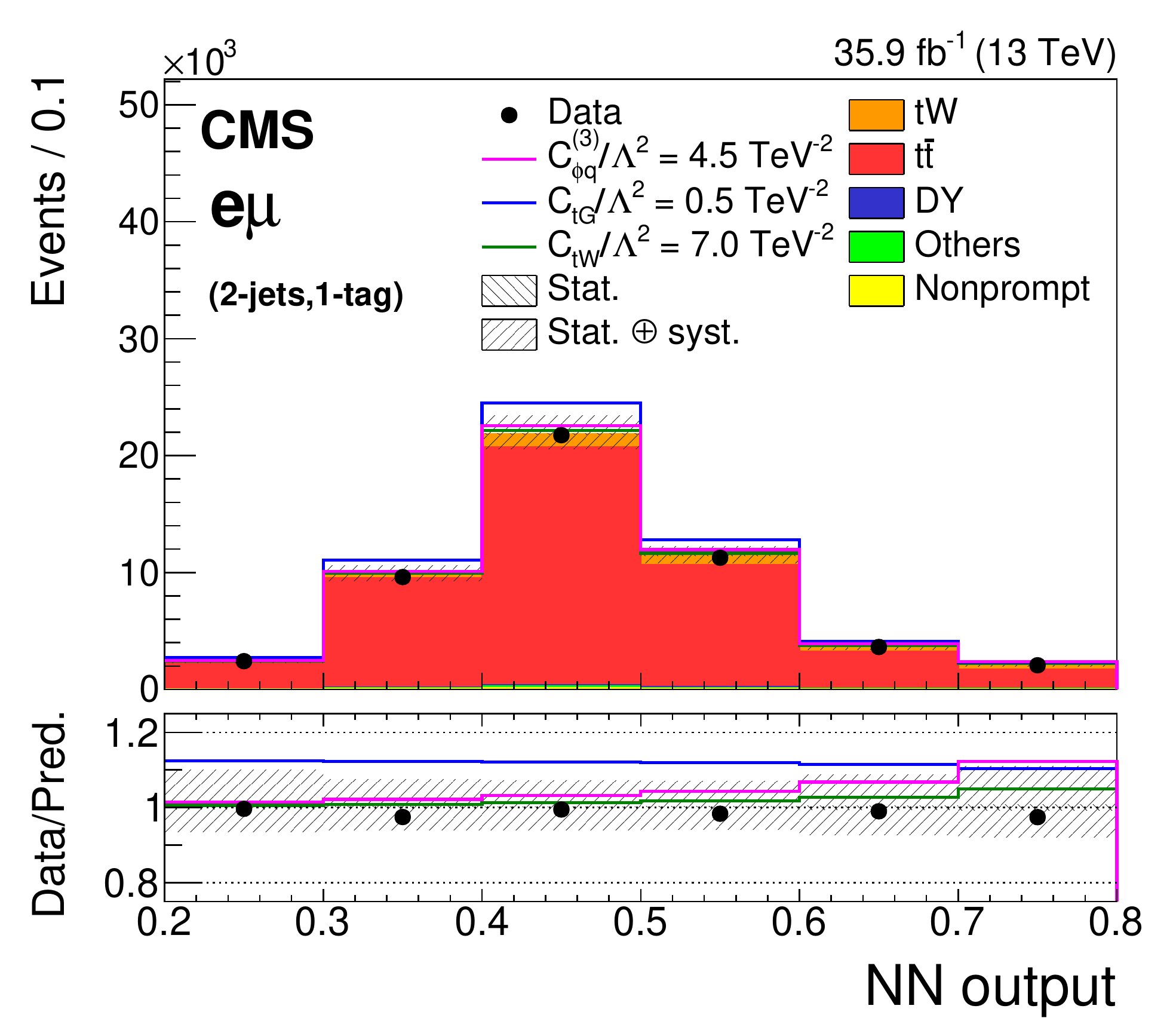}
      \includegraphics[width=0.45\textwidth]{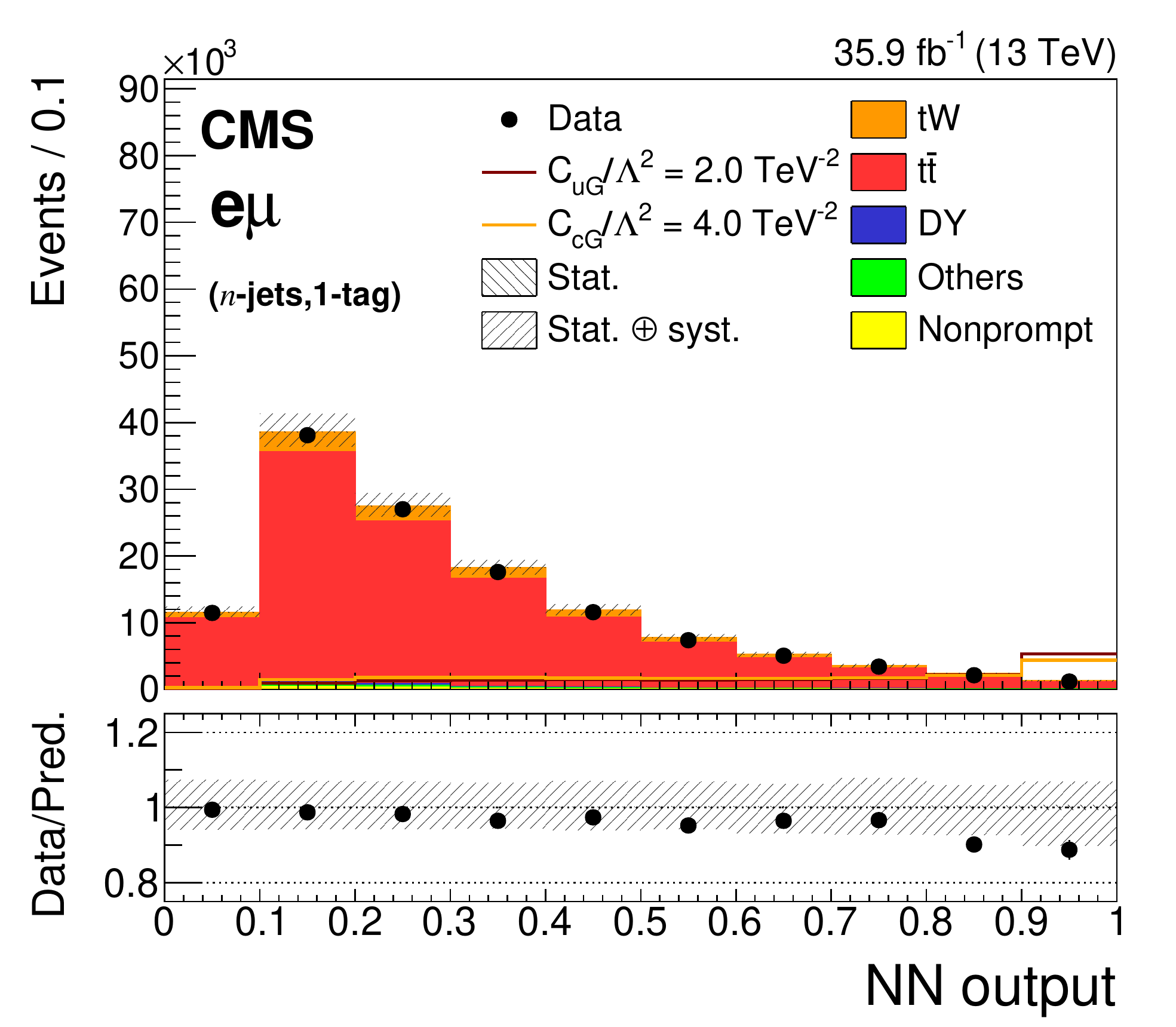}
      \includegraphics[width=0.45\textwidth]{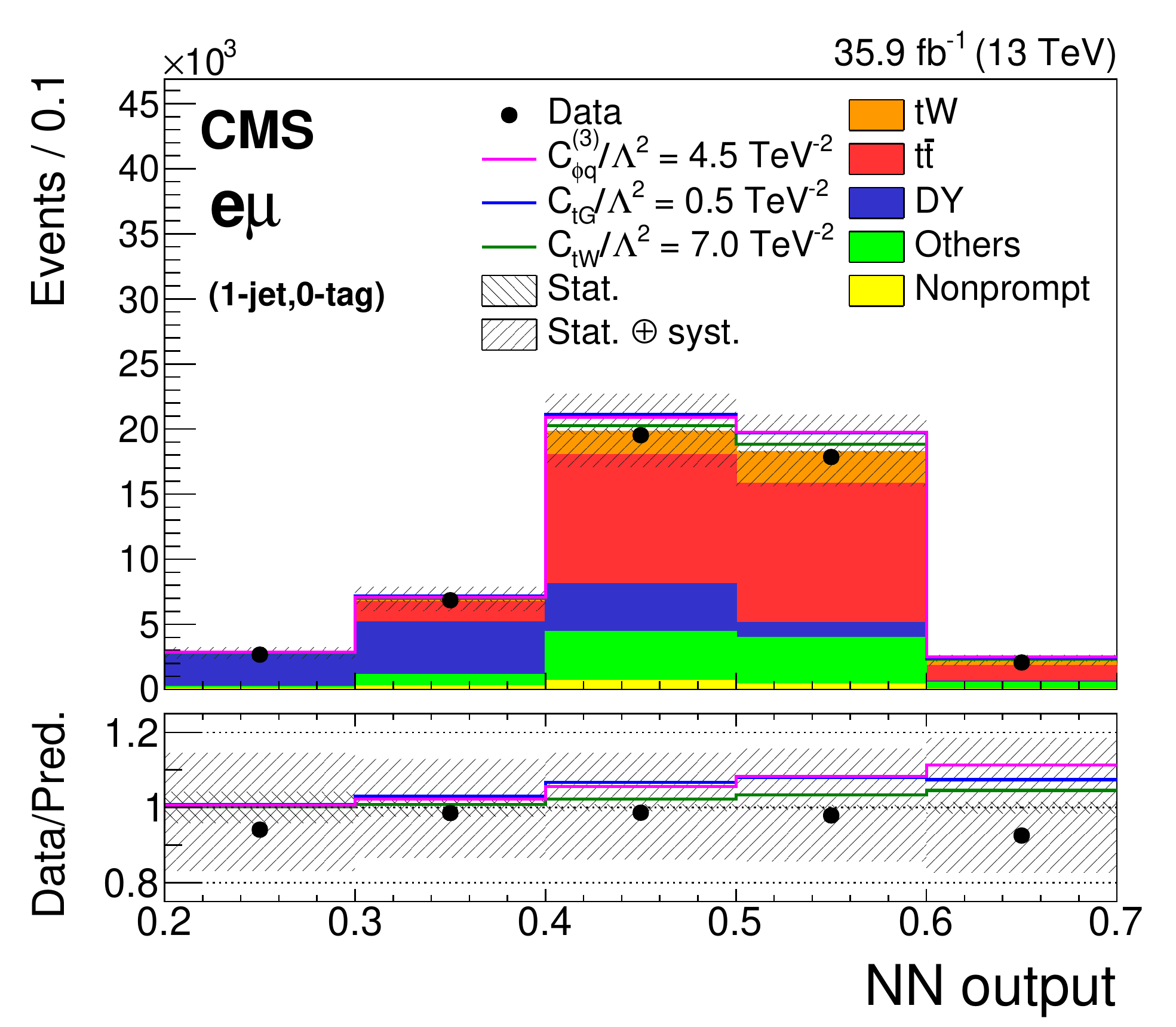}
      \includegraphics[width=0.45\textwidth]{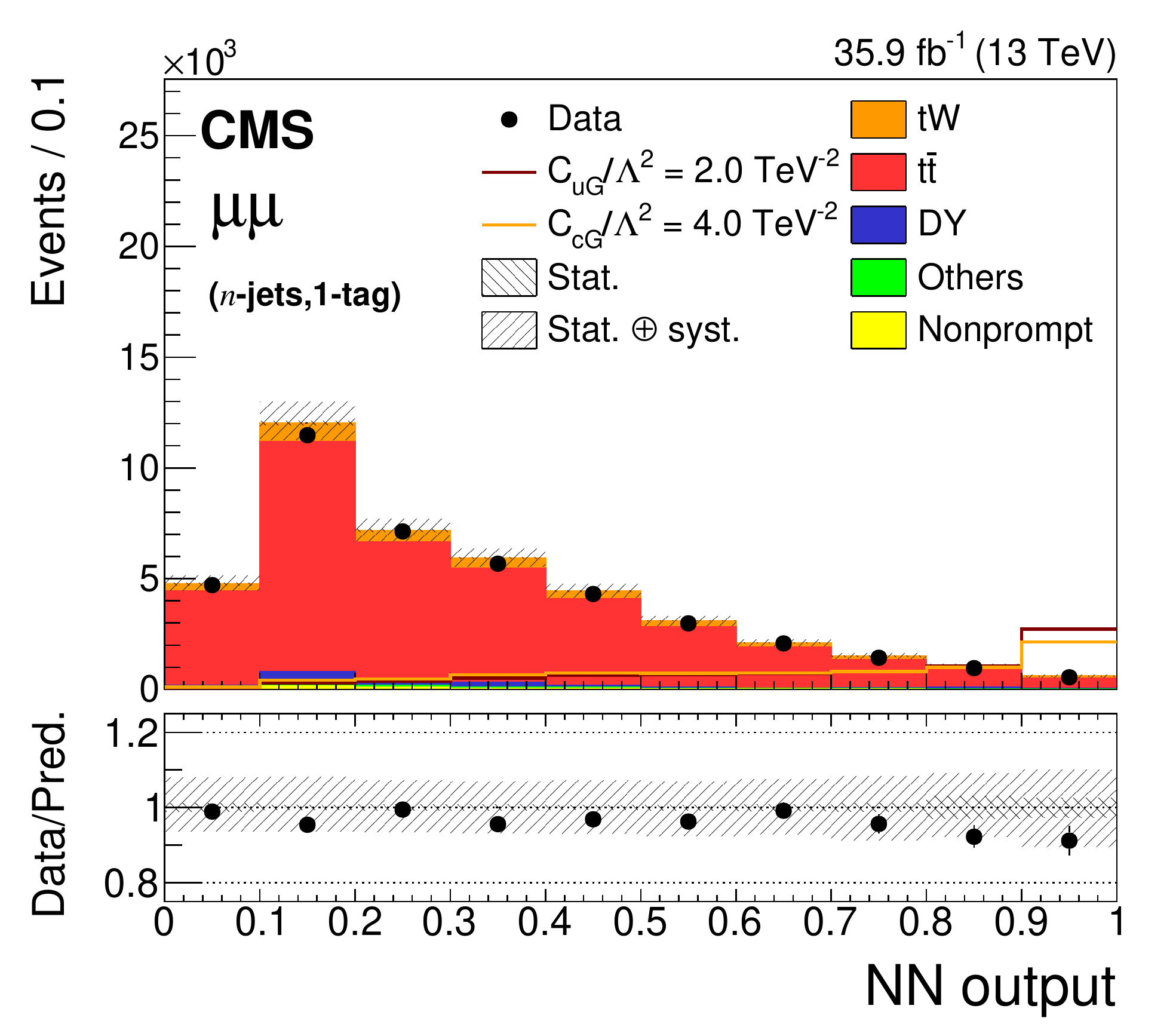}
    \caption{The NN output distributions for (left) data and simulation for the $\Pe\Pgm$ channel in 1-jet, 1-tag (upper) and 2-jets, 1-tag (middle) and 1-jet, 0-tag (lower) categories; and
for (right) data, simulation, and FCNC signals in the $n$-jets, 1-tag category used in the limit setting for the $\Pe\Pe$ (upper), $\Pe\Pgm$ (middle), and $\Pgm\Pgm$ (lower) channels.
The hatched bands correspond to the quadratic sum of the statistical and systematic uncertainties in the event yield for the sum of signal and background predictions. The ratios of data to the sum of the predicted yields are shown at the the lower panel of each graph. The narrow hatched bands represent the contribution from the statistical uncertainty in the MC simulation. In each plot, the expected distributions assuming specific values for the effective couplings (given in the legend) are shown as the solid curves.
    \label{fig:limit_binee2}}
\end{figure*}

The deviation from the SM $\cPqt\PW$ production because of the interference between the
SM and the $O_{\cPqt\cPG}$, $O_{\phi \cPq}^{(3)}$, and $O_{\cPqt\PW}$ operators is of the order of $1/\Lambda^2$.
It is assumed that the new physics scale $\Lambda$ is larger  than the scale we  probe. Therefore, $1/\Lambda^4$
contributions from the new physics terms are small compared to the contribution from the interference term.
The operator $O_{\phi \cPq}^{(3)}$ is similar to the SM $\PW\cPqt\cPqb$ operator and leads to a rescaling of the SM $\PW\cPqt\cPqb$ vertex~\cite{Zhang:2010dr}.
The $O_{\cPqt\PW}$ and $O_{\cPqt\cPG}$ operators lead to the right-handed Wtb interaction and a tensor-like ttg interaction, respectively, which are absent in the SM at the first order. 
Their effects have been investigated via the various kinematic distributions of the final-state particles considered in this analysis and are found to be not distinguishable from the SM $\cPqt\PW$ and \ttbar processes for unconstrained values of the effective couplings within the current precision on data.
After the selection described in Section 3, the dominant background comes from \ttbar production, with a contribution of about 90\%.
In order to observe deviations from SM $\cPqt\PW$ production in the presence of the $O_{\phi \cPq}^{(3)}$, $O_{\cPqt\PW}$, and $O_{\cPqt\cPG}$ effective operators, we need to separate $\cPqt\PW$ events from the large number of \ttbar events.
 Two independent NNs are trained to separate \ttbar events (the background) and $\cPqt\PW$ events (considered as the signal) in the (1-jet,1-tag) (NN$_{11}$) and (2-jets,1-tag) (NN$_{21}$) categories, which have significant signal contributions~\cite{Sirunyan:2018lcp}. For the $\Pe\Pgm$ channel, another NN is used for the   (1-jet,0-tag) (NN$_{10}$) category, in which the \ttbar, $\PW\PW$, and DY events are combined and are considered as the background.
A comparison between the observed data and the SM background prediction of the NN output shape in various ($n$-jets,$m$-tags) categories is shown for the $\Pe\Pe$ and $\Pgm\Pgm$ channels  in Fig. \ref{fig:limit_binee1} and for the $\Pe\Pgm$ channel in Fig. \ref{fig:limit_binee2} (left column).

The presence of the $O_{\cPqu\cPG}$ and $O_{\cPqc\cPG}$ operators changes the initial-state particle (see Fig. \ref{fig-feyn}), and leads to different kinematic distributions for the final-state particles, compared to the SM $\cPqt\PW$ process.
For these FCNC operators, new physics effects on final-state particle distributions are expected to be distinguishable from SM processes.
In order to search for new physics due to the $O_{\cPqu\cPG}$ and $O_{\cPqc\cPG}$ effective operators, an NN (NN$_{\mathrm{FCNC}}$) is used to separate SM backgrounds (\ttbar and $\cPqt\PW$ events together) and  new physics signals for events with exactly one {\cPqb}-tagged jet with no requirement on the number of light-flavor jets ($n$-jets,1-tag).
The comparison of the NN output for data, SM background  and signal ($\cPqt\PW$ events via FCNC interactions)  is   shown in Fig.~\ref{fig:limit_binee2} (right column) for the $\Pe\Pe$, $\Pe\Pgm$, and $\Pgm\Pgm$ channels.

The various input variables for training the NN introduced above are described below and are shown in Table ~\ref{tab:MVA_var}.
\begin{itemize}
  \item M$_{\ell \ell}$ (where $\ell$ = \Pe ~or \Pgm), invariant mass of dilepton system;
  \item $\pt^{\ell \ell}$, \pt of dilepton system;
  \item $\Delta \pt(\ell_1,\, \ell_2)$, $\pt^{\text{leading lepton}} - \pt^{\text{sub-leading lepton}}$;
  \item $\pt^{\ell_1}$, \pt of leading lepton;
  \item Centrality($\ell_1$, jet$_1$), scalar sum of \pt of the leading lepton and leading jet, over total energy of selected leptons and jets;
  \item Centrality($\ell_1,\, \ell_2$), scalar sum of \pt of the leading and sub-leading leptons, over total energy of selected leptons and jets;
  \item $\Delta \Phi(\ell \ell$, jet$_1$), $\Delta \Phi$ between dilepton system and leading jet where $\Phi$ is azimuthal angle;
  \item \pt($\ell \ell$, jet$_1$), \pt of dilepton and leading jet system;
  \item \pt($ \ell_1$, jet$_1$), \pt of leading lepton and leading jet system;
  \item Centrality($\ell\ell$, jet$_1$), scalar sum of \pt of the dilepton system and leading jet, over total energy of selected leptons and jets;
  \item $\Delta$R($\ell_1,\, \ell_2$), $\sqrt{\smash[b]{ \, ( \eta^{\ell_1} - \eta^{\ell_2})^2 \, + \, (\Phi^{\ell_1} - \Phi^{\ell_2})^2}}$;
  \item $\Delta$R($\ell_1$, jet$_1$), $\sqrt{\smash[b]{ \, ( \eta^{\ell_1} - \eta^{\text{jet}_1})^2 \, + \, (\Phi^{\ell_1} - \Phi^{ \text{jet}_1})^2}}$;
  \item M($\ell_1$, jet$_1$), invariant mass of leading lepton and leading jet;
  \item M(jet$_1$, jet$_2$), invariant mass of leading jet and sub-leading jet;
  \item $\Delta$R($\ell_1$, jet$_2$), $\sqrt{\smash[b]{ \, ( \eta^{\ell_1} - \eta^{\text{jet}_2})^2 \, + \, (\Phi^{\ell_1} - \Phi^{\text{jet}_2})^2}}$;
  \item $\Delta$R($\ell \ell$, jet$_1$), $\sqrt{\smash[b]{ \, ( \eta^{\ell \ell} - \eta^{\text{jet}_1})^2 \, + \, (\Phi^{\ell \ell} - \Phi^{\text{jet}_1})^2}}$;
  \item $\Delta \pt(\ell_2,~ \text{jet}_2)$, $\pt^{\ell_2} - \pt^{\text{jet}_2}$;
  \item M($\ell_2$, jet$_1$), invariant mass of sub-leading lepton and leading jet.
\end{itemize}

\begin{table}[!htb]
\topcaption{Input variables for the NN used in the analysis in various  bins of $n$-jets and $m$-tags. The symbols "$\times$" indicate the input variables
used in the four NNs.}
\centering
\label{tab:MVA_var}
\begin{tabular}{lcccc}
\hline
Variable                                                                           & NN$_{10}$             &NN$_{11}$                 & NN$_{21}$    &   NN$_{\text{FCNC}}$ \\ \hline
M$_{\ell \ell}$                                                                    & $\times$               &                          &              & $\times$              \\
$\pt^{\ell \ell }$                                                    & $\times$               &                          & $\times$      & $\times$              \\
$\Delta \pt(\ell_1,\, \ell_2)$& $\times$               &                          &              & $\times$              \\
$\pt^{\ell_1}$& $\times$               &                          & $\times$      & $\times$              \\
Centrality($\ell_1$, jet$_1$)&  $\times$               &                          &              & $\times$              \\
Centrality($\ell_1,\, \ell_2$)& $\times$               &                          &              & $\times$              \\
$\Delta \Phi(\ell \ell$, jet$_1$)& $\times$               & $\times$                  & $\times$      &                      \\
\pt($\ell \ell$, jet$_1$)&                       & $\times$                  &              & $\times$              \\
\pt($ \ell_1$, jet$_1$)&                       & $\times$                  &              &                      \\
Centrality($\ell\ell$, jet$_1$)&                       & $\times$                  &              &                      \\
$\Delta$R($\ell_1,\, \ell_2$)&                       & $\times$                  &              &                      \\
$\Delta$R($\ell_1$, jet$_1$)&                       & $\times$                  &              &                      \\
M($\ell_1$, jet$_1$)&                       &                          & $\times$      &                      \\
M(jet$_1$, jet$_2$)&                       &                          & $\times$      &                      \\
$\Delta$R($\ell_1$, jet$_2$)&                       &                          & $\times$      &                      \\
$\Delta$R($\ell \ell$, jet$_1$)&                       &                          & $\times$      & $\times$              \\
$\Delta \pt(\ell_2$, jet$_2$)&                       &                          & $\times$      &                      \\
M($\ell_2$, jet$_1$)&                       &                          &              & $\times$              \\  \hline
\end{tabular}
\end{table}

\section{Systematic uncertainties}
\label{sys}
The normalization and shape of the signal and the backgrounds are both affected by different sources of systematic uncertainty.
For  each  source,  an  induced  variation  can  be
parametrized, and treated as a nuisance parameter in the fit that is described in the next section.

A systematic uncertainty of 2.5\% is assigned to the integrated luminosity and is used for signal and background rates \cite{CMS-PAS-LUM-17-001}.
The efficiency corrections for trigger and offline selection  of leptons were estimated by comparing the efficiency measured in data and in MC simulation  using $\PZ\to\ell\ell$ events, based on a ``tag-and-probe'' method as in Ref.~\cite{Khachatryan:2010xn}. The scale factors obtained are varied by one standard deviation to take into account the corresponding uncertainties in the efficiency.
The jet energy scale and resolution uncertainties depend on \pt and $\eta$ of the jet and are computed by shifting the energy of each jet and propagating the variation to \ptvecmiss coherently~\cite{Chatrchyan:2011ds}.

The uncertainty in the \cPqb\ tagging is estimated by varying the \cPqb\ tagging scale factors within one standard deviation~\cite{Sirunyan:2017ezt}.
Effects of the uncertainty in the distribution of the number of pileup interactions are
evaluated by varying the effective inelastic $\Pp\Pp$ cross section used to predict the number of pileup interactions in MC simulation
by $\pm$4.6\% of its nominal value~\cite{Sirunyan:2018nqx}.

The uncertainty in the DY contribution in categories with one or two {\cPqb}-tagged jets is  considered to be  50 and 30\%  in the $\Pe\Pgm$ and same-flavor dilepton channels, respectively \cite{Khachatryan:2016kzg,Sirunyan:2018lcp}. For the DY normalization in the (1-jet,0-tag) category, an uncertainty of 15\% is assigned \cite{CMS:2015hfr}. In addition, systematic uncertainties related to the PDF, and to the renormalization and factorization scale uncertainty are taken into account for DY process in the (1-jet,0-tag) category.
The uncertainty in the yield of nonprompt lepton backgrounds is considered to be 50\% \cite{Sirunyan:2018lcp}.
Contributions to the background from \ttbar production in association with a boson, as well as diboson production, are estimated from simulation and a systematic uncertainty of 50\% is conservatively assigned \cite{Khachatryan:2014ewa}.

Various uncertainties originate from the theoretical predictions. The effect of the renormalization and factorization scale uncertainty from the \ttbar and $\cPqt\PW$ MC generators is estimated by varying the scales used during the generation of the simulation sample independently by a factor 0.5, 1 or 2. Unphysical cases, where one scale fluctuates up while the other fluctuates down, are not considered.
The top quark \pt reweighting procedure, discussed in Sec. \ref{cms}, is applied on top of the nominal \POWHEG prediction at NLO to account for the higher-order corrections.

The uncertainty in the PDFs for each simulated signal process is obtained using the replicas of the NNPDF 3.0 set~\cite{Butterworth:2015oua}.
The most recent measurement of the top quark mass by CMS yields a total uncertainty of $\pm 0.49$\GeV \cite{Khachatryan:2015hba}. We consider variations of the top quark mass due to this uncertainty and they are found to be insignificant.
At NLO QCD, $\cPqt\PW$ production is expected to interfere with \ttbar production \cite{Frixione:2008yi}. 
Two schemes for defining the $\cPqt\PW$ signal in a way that distinguishes it from the \ttbar production have been compared in the analysis: the ``diagram removal'' (DR), in which all
doubly resonant NLO $\cPqt\PW$ diagrams are removed,  and the ``diagram subtraction'' (DS), where a gauge-invariant subtractive term modifies the NLO $\cPqt\PW$ cross section to locally cancel the contribution from \ttbar production~\cite{Campbell:2005bb,Frixione:2008yi,Belyaev:2000me}. 
The DR method is used for the nominal $\cPqt\PW$ sample and the difference with respect to the sample simulated using the DS method  is taken as a systematic uncertainty.
The model parameter $h_{\text{damp}}$ in \ttbar
\POWHEG \cite{POWHEG} that controls the matching of the matrix elements to the \PYTHIA parton showers is varied from a top quark mass  default value of 172.5\GeV
by factors of 0.5 and 2 for estimating the uncertainties from the matching between jets from matrix element calculations and parton shower emissions.
The renormalization scale for QCD emissions in the initial-  and final-state radiation (ISR and FSR) is varied up and down by factors of 2 and $\sqrt{2}$, respectively, to account for parton shower QCD scale variation error in both \ttbar and $\cPqt\PW$ samples~\cite{Skands:2014pea}. In addition, several dedicated \ttbar samples are used to estimate shower modeling uncertainties in both underlying event and color re-connections \cite {Khachatryan:2016kzg,CMS-PAS-TOP-16-021,Skands:2014pea}. 
To estimate model uncertainties, $\cPqt\PW$ and \ttbar samples are generated with \POWHEG as described in Sec. \ref{cms}, varying the relevant model parameters with respect to the nominal samples.

\section{Constraints on the effective couplings}
\label{result}
The six Wilson coefficients sensitive to new physics contributions in top quark interactions, as defined in Eqs.~\ref{eq1}--\ref{eq5},  are tested separately in the observed data.
The event yields and the NN output distributions in each analysis category, summarized in Table \ref{tab:NNreg}, are used  to construct a binned likelihood function.
All sources of systematic uncertainty, described in Section \ref{sys}, are taken into account as nuisance parameters in the fit.
A simultaneous binned maximum-likelihood fit is performed to find the best fit value for each Wilson coefficient together with 68 and 95\% confidence intervals (CIs)~\cite{Khachatryan:2014jba}.
In this Section, distributions of the log-likelihood functions are shown with one nonzero effective coupling at a time for $\Lambda$ = 1\TeV.

The SM cross section prediction for the \ttbar and $\cPqt\PW$ processes, $\sigma_i^{(1)}$ and $\sigma_i^{(2)}$ (see Table \ref{xsVScoupling}), are accompanied by uncertainties in scales and PDFs.
These theoretical uncertainties can affect the bounds on the Wilson coefficients. In order to study this effect, the fit is performed on data,  while theoretical uncertainties are varied within one standard deviation and are shown together with the nominal results in the likelihood scan plots in Fig. \ref{LS}.
The nominal theoretical cross sections for \ttbar and $\cPqt\PW$ processes are varied by [+4.8\%,$-5.5\%$] and [+5.4\%,$-5.4\%$] , respectively.
These variations cover the uncertainties arising from the variations of factorization and renormalization scales and PDFs~\cite{Khachatryan:2016kzg,Kidonakis:2015nna}.
The scale variations for $\sigma_i^{(1)}$ and $\sigma_i^{(2)}$ are evaluated to be within 1 to 25\%. We assumed that the scale uncertainty is 100\% correlated among the terms $\sigma_{\mathrm {SM}}$,  $\sigma_i^{(1)}$, and $\sigma_i^{(2)}$.

\subsection {Exclusion limits on the \texorpdfstring{C$_{\cPG}$ effective coupling}{}}

In order to constrain the C$_{\cPG}$ coupling, the effect on the \ttbar rate in various ($n$-jets,$m$-tags) categories is considered.
The impact of the difference between the kinematic distributions of \ttbar events from the $O_{\cPG}$ interaction and from the SM interactions on the acceptance is evaluated to be 3\% for C$_{\cPG}\sim 1$. 
This uncertainty is considered only for the C$_{\cPG}$ coupling since the top \pt~spectrum is affected considerably by this operator, while other operators lead to a \pt~spectrum similar to the SM prediction for unconstrained values of the probed Wilson coefficients.
The fit is  performed simultaneously on the observed event yields in the categories presented in Fig.~\ref{fig:binee} in the (1-jet,1-tag), (2-jets,1-tag), and ($\geq$2-jets,2-tags) categories for the $\Pe\Pe$, $\Pe\Pgm$, and $\Pgm\Pgm$ channels.
In addition, the (1-jet,0-tag) category is included only for the $\Pe\Pgm$ channel.
The main limiting factor on the constraints in the C$_{\cPG}$ coupling is the uncertainty in the signal acceptance found after maximizing the likelihood, followed by uncertainties in the integrated luminosity calibration and the trigger scale factor.

The results of the fit for the individual channels and for all channels combined are listed in the first row of Table \ref{figLIM}.

\begin{figure*}[htbp]
  \centering
      \includegraphics[width=0.45\textwidth]{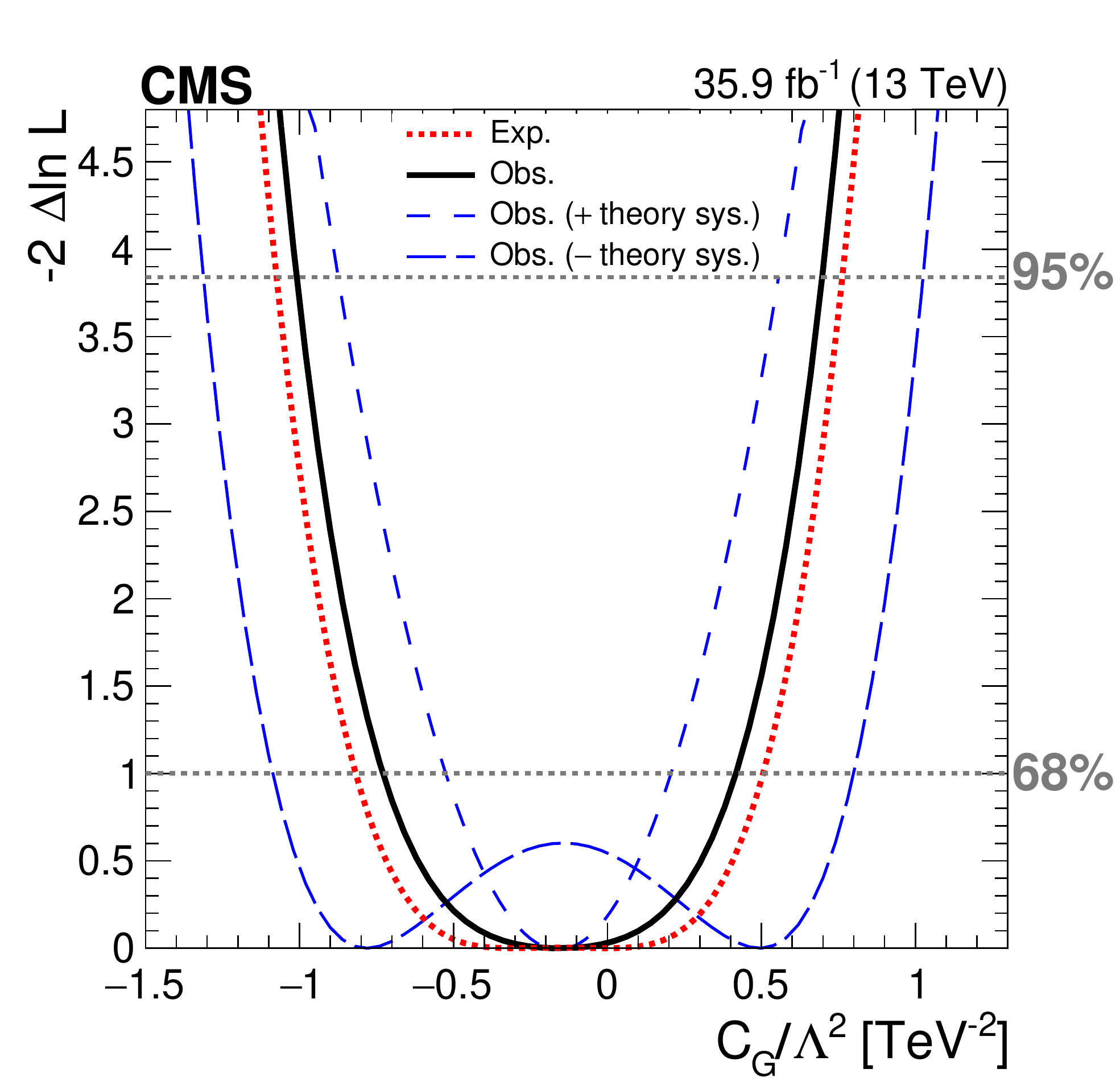}
      \includegraphics[width=0.45\textwidth]{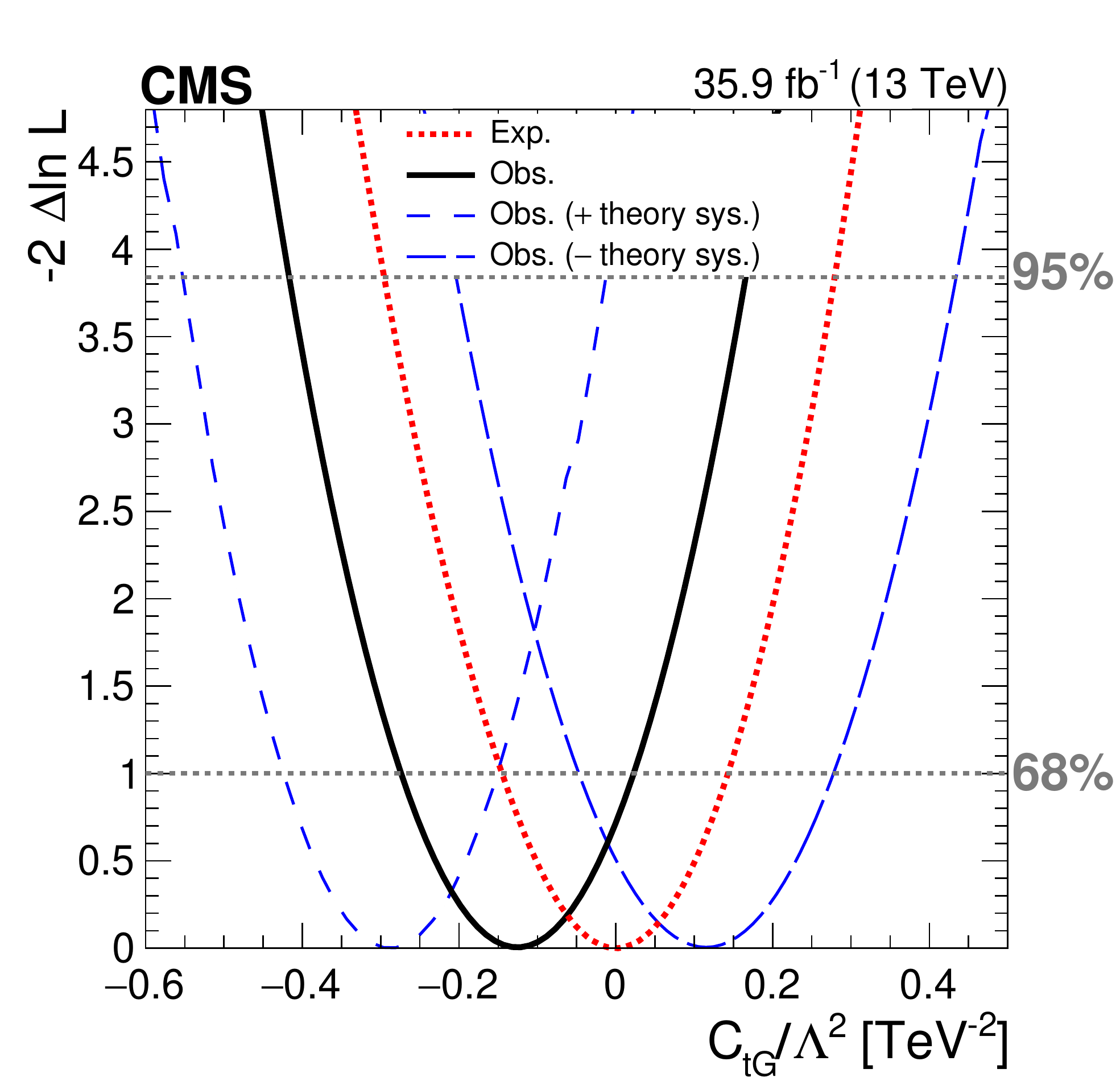}
      \includegraphics[width=0.45\textwidth]{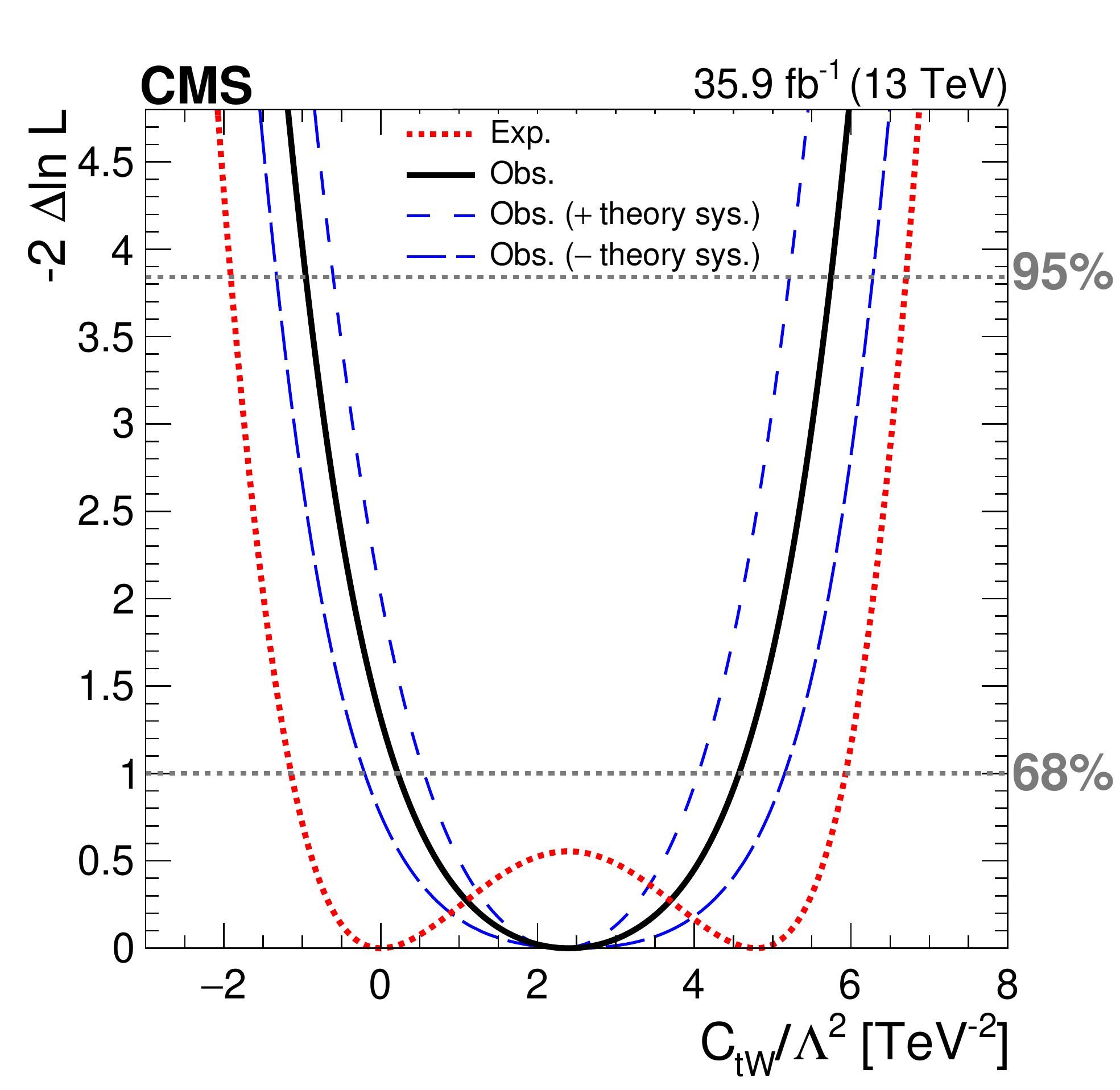}
      \includegraphics[width=0.45\textwidth]{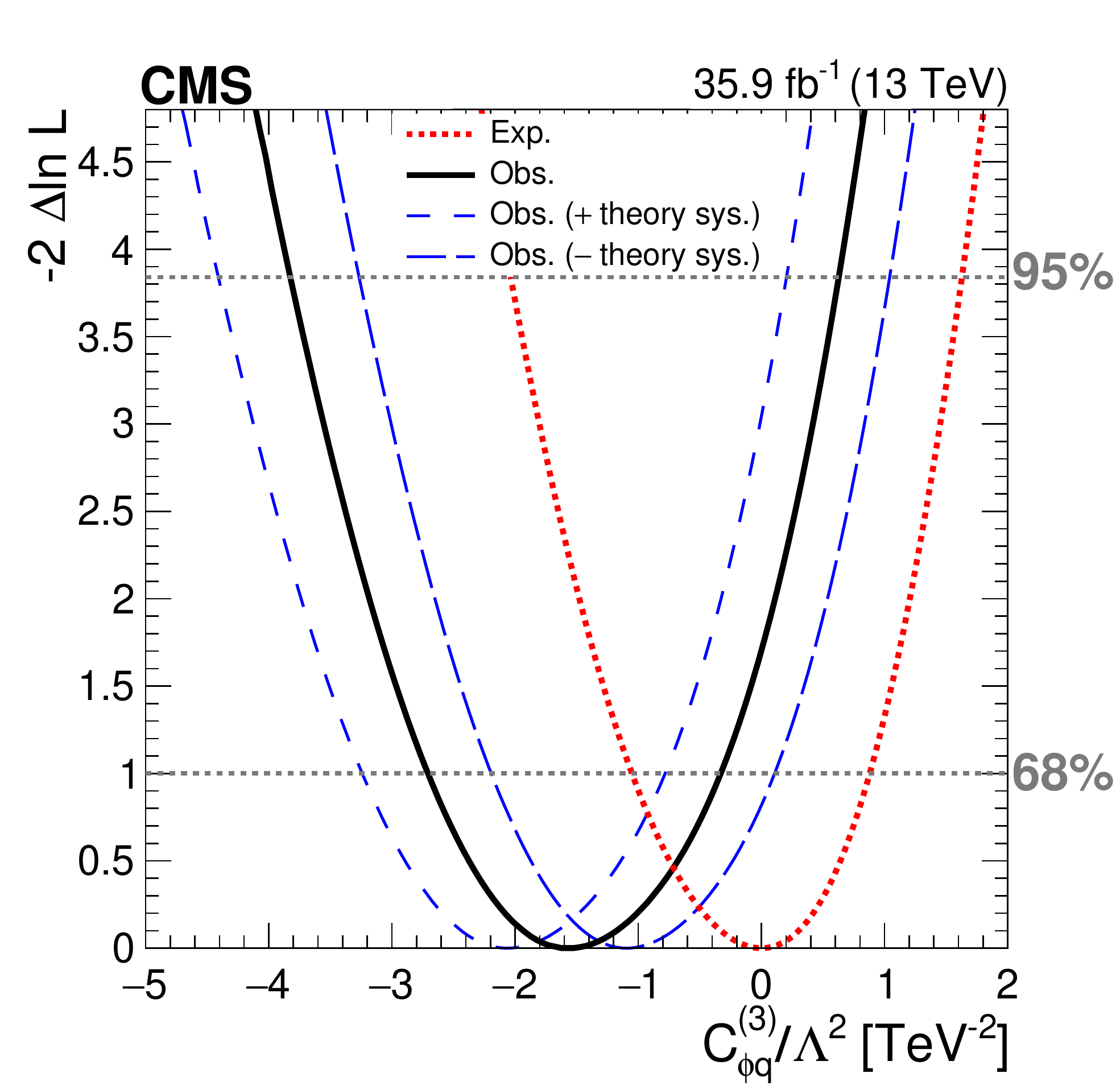}
      \includegraphics[width=0.45\textwidth]{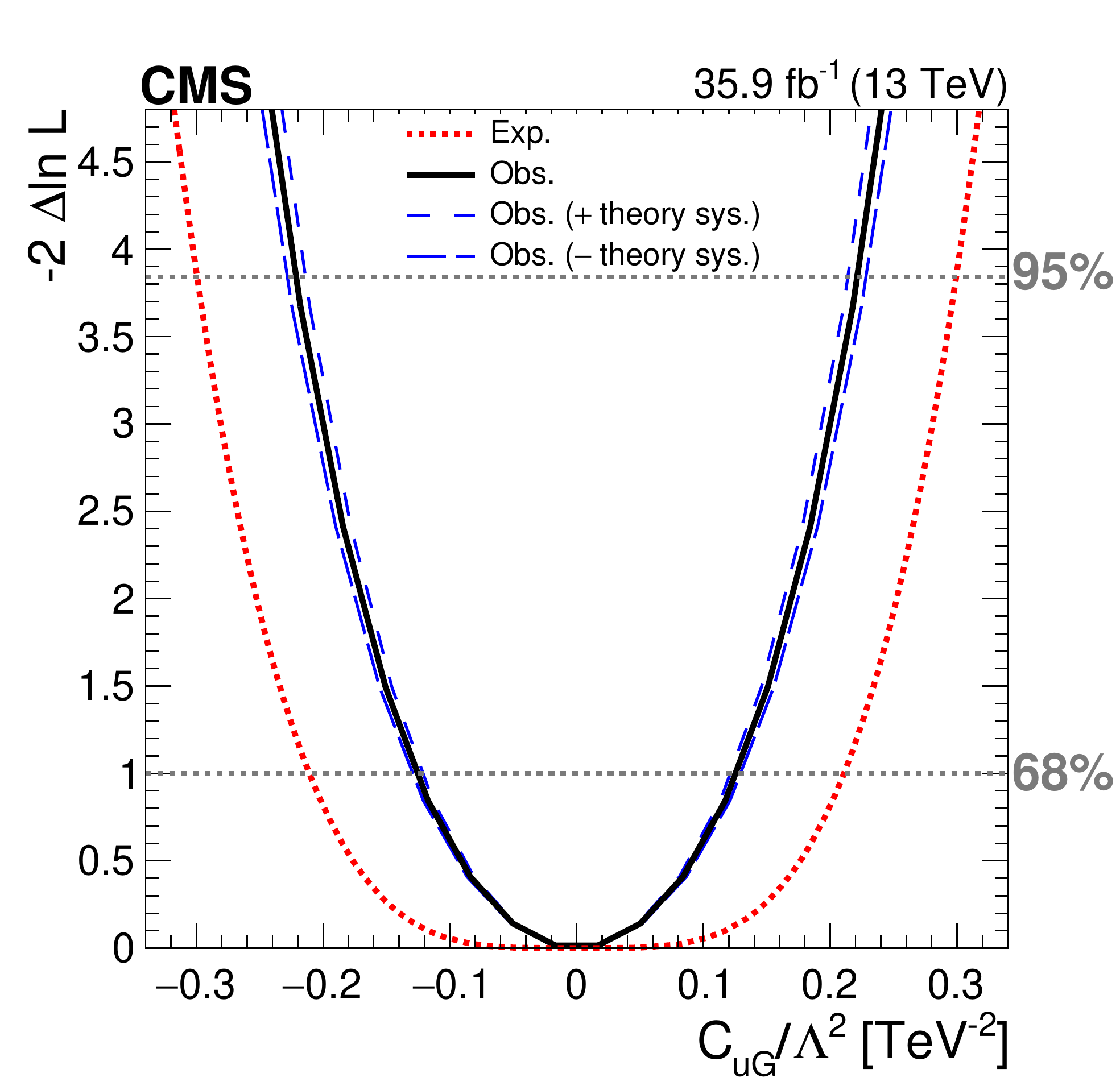}
      \includegraphics[width=0.45\textwidth]{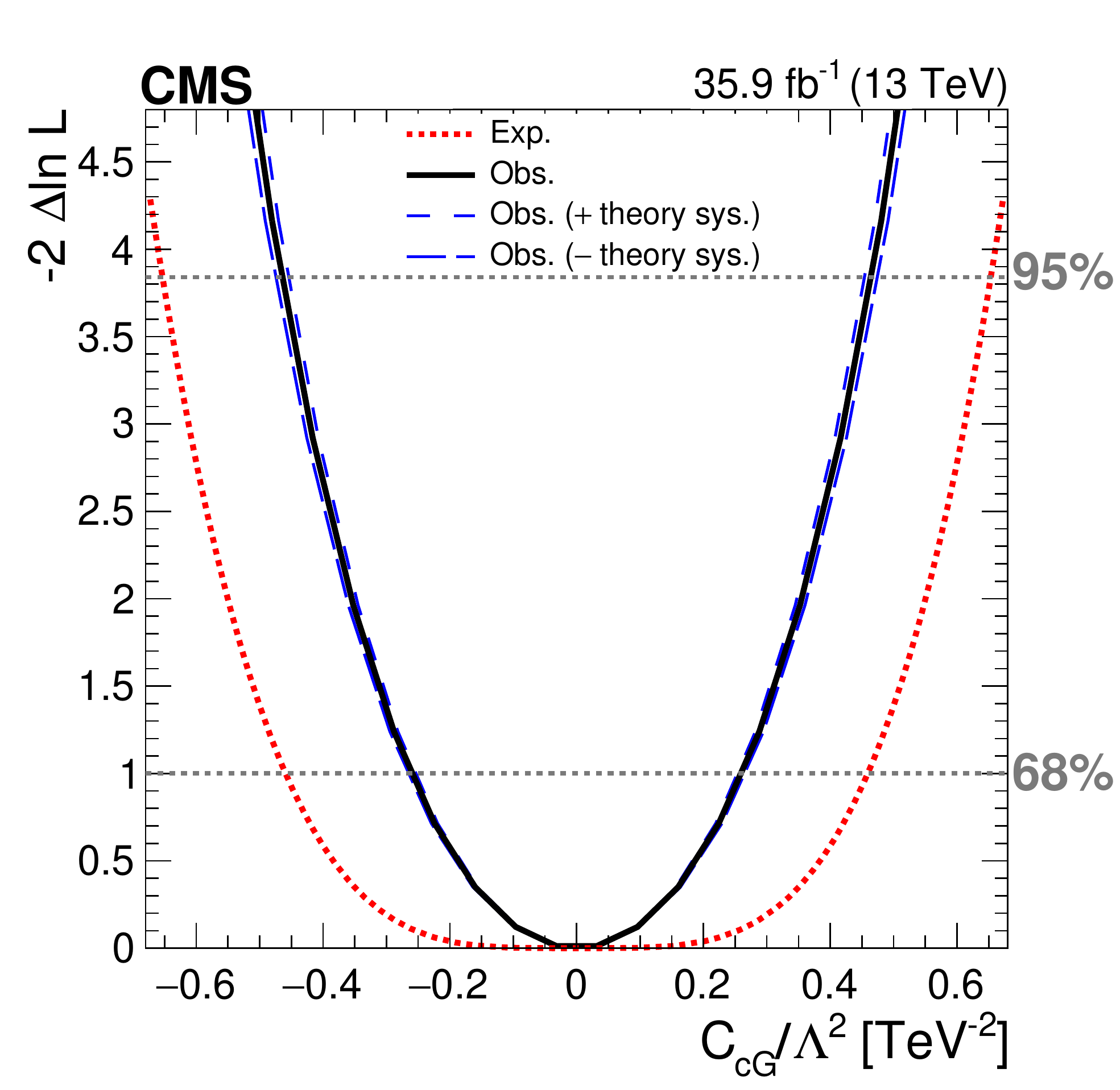}
    \caption{Observed (solid) and expected (dotted) log likelihoods for  the effective couplings: C$_{\cPG}$ (upper left), C$_{\cPqt\cPG}$ (upper right), C$_{\cPqt\PW}$ (middle left), C$_{\phi \cPq}$ (middle right), C$_{\cPqu\cPG}$ (lower left), and C$_{\cPqc\cPG}$ (lower right). The dashed curves represent fits to the observed data with the variations of normalization due to the theoretical uncertainties.
    \label{LS}}
\end{figure*}

The results of the likelihood scans of the C$_{\cPG}$ coupling are shown in Fig. \ref{LS} (upper left plot).
The likelihood scan result of the nominal fit, in which the nominal values of $\sigma_{\mathrm {SM}}$,  $\sigma_i^{(1)}$, and $\sigma_i^{(2)}$ terms are assumed, is shown as the thick curve.
The thin dashed curves are the results of the fit to the observed data when the assumed values of the  $\sigma_{\mathrm {SM}}$,  $\sigma_i^{(1)}$, and $\sigma_i^{(2)}$ terms are varied due to the scale and PDF uncertainties.
As a second-order parametrization, given by Eq.~\ref{eq2x} is used to fit the data, the resulting likelihood function could have two minima, as can be seen in some of the plots in Fig. \ref{LS}.

\subsection {Exclusion limits on the \texorpdfstring{C$_{\cPqt\cPG}$, C$_{\phi \cPq}^{(3)}$, and C$_{\cPqt\PW}$ effective couplings}{}}
In order to set limits on the effective couplings C$_{\cPqt\cPG}$, C$_{\phi \cPq}^{(3)}$, and C$_{\cPqt\PW}$, we utilize the NN output distributions for both data and MC expectation in the (1-jet,1-tag)
and (2-jets,1-tag) regions and event yields in the ($\geq$2-jets,2-tags)
region for the three dilepton channels. The inclusion of the ($\geq$2-jets,2-tags) and (2-jets,1-tag) categories provides a constraint of the normalization and systematic uncertainties in the \ttbar background.
In addition, the (1-jet,0-tag) category is included for the $\Pe\Pgm$ channel to increase the signal sensitivity.
The results of the likelihood scans of the C$_{\cPqt\cPG}$, C$_{\phi \cPq}^{(3)}$, and C$_{\cPqt\PW}$ Wilson coefficients are shown in Fig. \ref{LS} for the combination of all channels.
The inclusion of the C$_{\cPqt\cPG}$ coupling to the tW process  tightens the 2 standard deviations band by 7\%.
The results for the individual channels, and the combined results are listed in Table \ref{figLIM} (second, third, and fourth rows).
The three main sources of uncertainty that affect the interval determination are uncertainties in the DY estimation, integrated luminosity, and lepton identification scale factors for C$_{\cPqt\cPG}$; jet energy scale, tt and tW interference at NLO, and statistical uncertainty in MC samples for C$_{\phi \cPq}^{(3)}$; statistical uncertainty in data, jet energy scale, and the \POWHEG matching method for C$_{\cPqt\PW}$ effective couplings.

\subsection {Exclusion limits on the \texorpdfstring{C$_{\cPqu\cPG}$ and C$_{\cPqc\cPG}$  effective couplings}{}}
Since the $\cPqt\PW$ production via FCNC interactions does not interfere with the SM $\cPqt\PW$ process (with the assumption of $\abs{\text{V}_{\cPqt\cPqd}} = \abs{\text{V}_{\cPqt\cPqs}} = 0$), the FCNC signal sample  is used  to set upper bounds on the related Wilson coefficients.
Events with exactly one {\cPqb}-tagged jet are included in the limit setting procedure with no requirement on the number of light-flavor jets ($n$-jets,1-tag).
The observed  (median expected) $95\%$ confidence level (\CL) upper limits on the product of cross section times branching fractions $\sigma(\Pp\Pp\to\cPqt\PW)\mathcal{B}(\PW\to\ell\cPgn)^2$ for the C$_{\cPqu\cPG}$ and  C$_{\cPqc\cPG}$ FCNC signals for the combination of the $\Pe\Pe$, $\Pgm\Pgm$, and $\Pe\Pgm$ channels  are found to be 0.11 (0.20)\unit{pb}  and 0.13 (0.26)\unit{pb}, respectively.
These results are used to calculate  upper limits on the Wilson coefficients C$_{\cPqu\cPG}$, C$_{\cPqc\cPG}$, and  on the branching fractions $\mathcal{B}(\cPqt\to\cPqu\Pg)$ and $\mathcal{B}(\cPqt\to\cPqc\Pg)$. The limits on the C$_{\cPqu\cPG}$ and  C$_{\cPqc\cPG}$ couplings are summarized in the last two rows of Table \ref{figLIM}, and correspond to the observed (expected) limits  on $\mathcal{B}(\cPqt\to\cPqu\Pg)<$ 0.12 (0.22)\% and $\mathcal{B}(\cPqt\to\cPqc\Pg)<$ 0.53 (1.05)\% at $95\%$ \CL.
The statistical uncertainty in data is the dominant source of uncertainty affecting the limits on the FCNC couplings. The second and third most important uncertainties originate from \ttbar and $\cPqt\PW$ interferences at NLO and FSR in \ttbar events. 

\begin{figure*}[htb!p]
  \centering
      \includegraphics[width=0.75\textwidth]{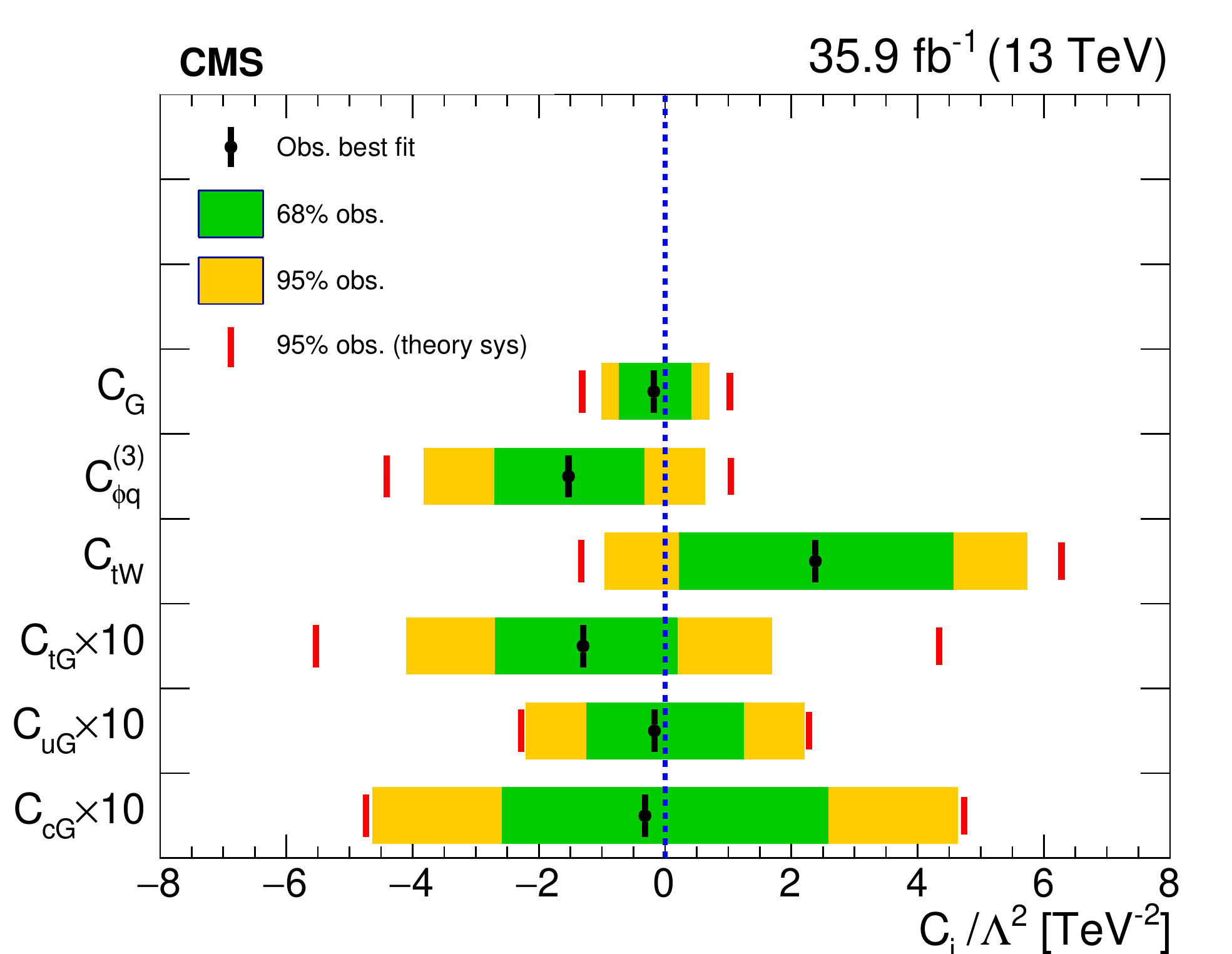}
    \caption{Observed best fits together with one and two standard deviation bounds on the top quark effective couplings. The dashed line shows the SM expectation and the vertical lines indicate the $95\%$ \CL bounds including the theoretical uncertainties. }
    \label{fin}
\end{figure*}

The observed best fit together with one and two standard deviation bounds on the six Wilson coefficients, C$_{\phi \cPq}^{(3)}$, C$_{\cPqt\PW}$, C$_{\cPqt\cPG}$, C$_{\cPG}$, C$_{\cPqu\cPG}$, and C$_{\cPqc\cPG}$, obtained from  the combination of all channels are shown in Fig. \ref{fin}. 
Table \ref{syseff} summarizes the effect of the most important uncertainty sources on the observed allowed intervals.

\begin{table*}[htb]
\centering
\topcaption{Summary of the observed and expected allowed intervals on the  effective couplings  obtained in the $\Pe\Pe$, $\Pe\Pgm$, and $\Pgm\Pgm$ channels, and all channels combined.
All sources of systematic uncertainty, described in Section 5, are taken into account with the exception of the uncertainties on the SM cross section predictions for the tt and tW processes.}
\cmsTable{
\begin{tabular}{llllllll}
\hline
Effective        & \multirow{2}{*}{Channel}                                    & \multicolumn{3}{c}{Observed [TeV$^{-2}$]}         & \multicolumn{3}{c}{Expected [TeV$^{-2}$]}   \\
coupling                     &                                            & Best fit  & [68\% CI] & [95\% CI]                   & Best fit  & [68\% CI] & [95\% CI]             \\ \hline
\multirow{4}{*}{C$_{\cPG} / \Lambda^2$}             & $\Pe\Pe$         & $-0.14  $&$ [-0.82 ,  0.51]$&$[-1.14 , 0.83]$     & $0.00 $&$ [-0.90 , 0.59]$&$[-1.20 , 0.88]$ \\
                              & $\Pe\Pgm$                            & $-0.18  $&$ [-0.73 ,  0.42]$&$[-1.01 , 0.70]$     & $0.00 $&$ [-0.82 , 0.51]$&$[-1.08 , 0.77]$ \\
                              & $\Pgm\Pgm$                          & $-0.14  $&$ [-0.75 ,  0.44]$&$[-1.06 , 0.75]$     & $0.00 $&$ [-0.88 , 0.57]$&$[-1.16 , 0.85]$ \\
                              & Combined                          & $-0.18  $&$ [-0.73 ,  0.42]$&$[-1.01 , 0.70]$     & $0.00 $&$ [-0.82 , 0.51]$&$[-1.07 , 0.76]$ \\ [\cmsTabSkip]
\multirow{4}{*}{C$_{\phi \cPq}^{(3)} / \Lambda^2$} & $\Pe\Pe$          & $1.12 $  &$ [-1.18 ,  2.89]$&$[-4.03 , 4.37]$     & $0.00 $&$ [-2.53 , 1.74]$&$[-6.40 , 3.27]$ \\
                              & $\Pe\Pgm$                            & $-0.70  $&$ [-2.16 ,  0.59]$&$[-3.74 , 1.61]$     & $0.00 $&$ [-1.34 , 1.12]$&$[-2.57 , 2.15]$ \\
                              & $\Pgm\Pgm$                          & $1.13 $  &$ [-0.87 ,  2.86]$&$[-3.58 , 4.46]$     & $0.00 $&$ [-2.20 , 1.92]$&$[-4.68 , 3.66]$ \\
                              & Combined                          & $-1.52  $&$ [-2.71 , -0.33]$&$[-3.82 , 0.63]$     & $0.00 $&$ [-1.05 , 0.88]$&$[-2.04 , 1.63]$ \\ [\cmsTabSkip]
\multirow{4}{*}{C$_{\cPqt\PW} / \Lambda^2$} & $\Pe\Pe$                    & $6.18 $  &$ [-3.02 ,  7.81]$&$[-4.16 , 8.95]$     & $0.00 $&$ [-2.02 , 6.81]$&$[-3.33 , 8.12]$ \\
                              & $\Pe\Pgm$                            & $1.64 $  &$ [-0.80 ,  5.59]$&$[-1.89 , 6.68]$     & $0.00 $&$ [-1.40 , 6.19]$&$[-2.39 , 7.18]$ \\
                              & $\Pgm\Pgm$                          & $-1.40  $&$ [-3.00 ,  7.79]$&$[-4.23 , 9.01]$     & $0.00 $&$ [-2.18 , 6.97]$&$[-3.63 , 8.42]$ \\
                              & Combined                          & $2.38 $  &$ [ 0.22 ,  4.57]$&$[-0.96 , 5.74]$     & $0.00 $&$ [-1.14 , 5.93]$&$[-1.91 , 6.70]$ \\ [\cmsTabSkip]
\multirow{4}{*}{C$_{\cPqt\cPG} / \Lambda^2$} & $\Pe\Pe$                    & $-0.19  $&$ [-0.40 ,  0.02]$&$[-0.65 , 0.22]$     & $0.00 $&$ [-0.22 , 0.21]$&$[-0.44 , 0.41]$ \\
                              & $\Pe\Pgm$                            & $-0.03  $&$ [-0.19 ,  0.11]$&$[-0.34 , 0.27]$     & $0.00 $&$ [-0.17 , 0.15]$&$[-0.34 , 0.29]$ \\
                              & $\Pgm\Pgm$                          & $-0.15  $&$ [-0.34 ,  0.02]$&$[-0.53 , 0.19]$     & $0.00 $&$ [-0.19 , 0.18]$&$[-0.40 , 0.35]$ \\
                              & Combined                          & $-0.13  $&$ [-0.27 ,  0.02]$&$[-0.41 , 0.17]$     & $0.00 $&$ [-0.15 , 0.14]$&$[-0.30 , 0.28]$ \\ [\cmsTabSkip]
\multirow{4}{*}{C$_{\cPqu\cPG} / \Lambda^2$} & $\Pe\Pe$                    & $-0.017 $&$ [-0.22 ,  0.22]$&$[-0.37 , 0.37]$     & $0.00 $&$ [-0.29 , 0.29]$&$[-0.42 , 0.42]$ \\
                              & $\Pe\Pgm$                            & $-0.017 $&$ [-0.17 ,  0.17]$&$[-0.29 , 0.29]$     & $0.00 $&$ [-0.26 , 0.26]$&$[-0.38 , 0.38]$ \\
                              & $\Pgm\Pgm$                          & $-0.017 $&$ [-0.17 ,  0.17]$&$[-0.29 , 0.29]$     & $0.00 $&$ [-0.27 , 0.27]$&$[-0.38 , 0.38]$ \\
                              & Combined                          & $-0.017 $&$ [-0.13 ,  0.13]$&$[-0.22 , 0.22]$     & $0.00 $&$ [-0.21 , 0.21]$&$[-0.30 , 0.30]$ \\ [\cmsTabSkip]
\multirow{4}{*}{C$_{\cPqc\cPG} / \Lambda^2$} & $\Pe\Pe$                    & $-0.032 $&$ [-0.47 ,  0.47]$&$[-0.78 , 0.78]$     & $0.00 $&$ [-0.63 , 0.63]$&$[-0.92 , 0.92]$ \\
                              & $\Pe\Pgm$                            & $-0.032 $&$ [-0.34 ,  0.34]$&$[-0.60 , 0.60]$     & $0.00 $&$ [-0.56 , 0.56]$&$[-0.81 , 0.81]$ \\
                              & $\Pgm\Pgm$                          & $-0.032 $&$ [-0.36 ,  0.36]$&$[-0.63 , 0.63]$     & $0.00 $&$ [-0.58 , 0.58]$&$[-0.84 , 0.84]$ \\
                              & Combined                          & $-0.032 $&$ [-0.26 ,  0.26]$&$[-0.46 , 0.46]$     & $0.00 $&$ [-0.46 , 0.46]$&$[-0.65 , 0.65]$ \\ \hline
\end{tabular}
}
\label{figLIM}
\end{table*}

\begin{table*}[htb]
\centering
\topcaption{ Estimation of the effect of the most important uncertainty sources on the observed allowed intervals of in the fit.}
\begin{tabular}{lllllll}
\hline
Uncertainty                    & C$_{\cPG}$ & C$_{\phi \cPq}^{(3)}$ & C$_{\cPqt\PW}$ & C$_{\cPqt\cPG}$ & C$_{\cPqu\cPG}$ & C$_{\cPqc\cPG}$ \\
\hline
Trigger                        & 10.2\%  & 2.3\%      & 7.0\%   & 2.9\%    & 1.7\%    & 2.5\%            \\
Lepton ident./isolation        & 7.4\%   & 1.1\%      & 1.2\%   & 23.0\%   & $<$1\%   & $<$1\%            \\
Jet energy scale               & $<$1\%  & 25.0\%     & 17.8\%  & 4.9\%    & $<$1\%   & $<$1\%            \\
$\cPqt\PW$ DS/DR               & $<$1\%  & 24.2\%     & 4.4\%   & 3.0\%    & 7.6\%    & 7.8\%            \\
ME/PS matching                 & $<$1\%  & 4.9\%      & 9.9\%   & 1.2\%    & $<$1\%   & $<$1\%            \\
ISR scale                      & $<$1\%  & 5.0\%      & 5.6\%   & $<$1\%   & $<$1\%   & $<$1\%            \\
FSR scale                      & 5.8\%   & 4.4\%      & 4.0\%   & 10.2\%   & $<$1\%   & $<$1\%            \\
DY background                  & $<$1\%  & 7.5\%      & 5.5\%   & 21.5\%   & $<$1\%   & $<$1\%            \\
Nonprompt background           & $<$1\%  & 1.4\%      & 5.8\%   & $<$1\%   & $<$1\%   & $<$1\%            \\
Integrated luminosity          & 13.1\%  & $<$1\%     & 1.1\%   & 18.8\%   & $<$1\%   & $<$1\%            \\
Statistical                    & 5.8\%   & 2.3\%      & 23.7\%  & $<$1\%   & 72.6\%   & 73.6\%            \\
MC statistical                 & $<$1\%  & 12.1\%     & 3.7\%   & 5.2\%    & 2.9\%    & 2.5\%            \\ 
\hline
\end{tabular}
\label{syseff}
\end{table*}

\section{Summary}
\label{summary}
A search for new physics in top quark interactions is performed using \ttbar and $\cPqt\PW$ events in dilepton final states.
The analysis is based on data collected in $\Pp\Pp$ collisions at 13\TeV by the CMS detector in 2016, corresponding to an integrated luminosity of 35.9\fbinv.
No  significant  excess  above  the  standard  model  background expectation is observed.
For the first time, both \ttbar and $\cPqt\PW$ production are used simultaneously in a model independent search for effective couplings.
The six effective couplings, C$_{\cPG}$, C$_{\cPqt\cPG}$, C$_{\cPqt\PW}$, C$_{\phi \cPq}^{(3)}$, C$_{\cPqu\cPG}$, and C$_{\cPqc\cPG}$ are constrained using a dedicated multivariate analysis.
The constraints presented, obtained by considering one operator at a time, are a useful first step toward more global approaches.

\clearpage

\begin{acknowledgments}
We congratulate our colleagues in the CERN accelerator departments for the excellent performance of the LHC and thank the technical and administrative staffs at CERN and at other CMS institutes for their contributions to the success of the CMS effort. In addition, we gratefully acknowledge the computing centres and personnel of the Worldwide LHC Computing Grid for delivering so effectively the computing infrastructure essential to our analyses. Finally, we acknowledge the enduring support for the construction and operation of the LHC and the CMS detector provided by the following funding agencies: BMWFW and FWF (Austria); FNRS and FWO (Belgium); CNPq, CAPES, FAPERJ, and FAPESP (Brazil); MES (Bulgaria); CERN; CAS, MoST, and NSFC (China); COLCIENCIAS (Colombia); MSES and CSF (Croatia); RPF (Cyprus); SENESCYT (Ecuador); MoER, ERC IUT, and ERDF (Estonia); Academy of Finland, MEC, and HIP (Finland); CEA and CNRS/IN2P3 (France); BMBF, DFG, and HGF (Germany); GSRT (Greece); OTKA and NIH (Hungary); DAE and DST (India); IPM (Iran); SFI (Ireland); INFN (Italy); MSIP and NRF (Republic of Korea); LAS (Lithuania); MOE and UM (Malaysia); BUAP, CINVESTAV, CONACYT, LNS, SEP, and UASLP-FAI (Mexico); MBIE (New Zealand); PAEC (Pakistan); MSHE and NSC (Poland); FCT (Portugal); JINR (Dubna); MON, RosAtom, RAS, RFBR and RAEP (Russia); MESTD (Serbia); SEIDI, CPAN, PCTI and FEDER (Spain); Swiss Funding Agencies (Switzerland); MST (Taipei); ThEPCenter, IPST, STAR, and NSTDA (Thailand); TUBITAK and TAEK (Turkey); NASU and SFFR (Ukraine); STFC (United Kingdom); DOE and NSF (USA).

\hyphenation{Rachada-pisek} Individuals have received support from the Marie-Curie program and the European Research Council and EPLANET (European Union); the Leventis Foundation; the A. P. Sloan Foundation; the Alexander von Humboldt Foundation; the Belgian Federal Science Policy Office; the Fonds pour la Formation \`a la Recherche dans l'Industrie et dans l'Agriculture (FRIA-Belgium); the Agentschap voor Innovatie door Wetenschap en Technologie (IWT-Belgium); the Ministry of Education, Youth and Sports (MEYS) of the Czech Republic; the Council of Science and Industrial Research, India; the HOMING PLUS program of Foundation for Polish Science, cofinanced from European Union, Regional Development Fund; the Compagnia di San Paolo (Torino); the Consorzio per la Fisica (Trieste); MIUR project 20108T4XTM (Italy); the Thalis and Aristeia programs cofinanced by EU-ESF and the Greek NSRF; and the National Priorities Research Program by Qatar National Research Fund.
\end{acknowledgments}

\bibliography{auto_generated}

\cleardoublepage \appendix\section{The CMS Collaboration \label{app:collab}}\begin{sloppypar}\hyphenpenalty=5000\widowpenalty=500\clubpenalty=5000\vskip\cmsinstskip
\textbf{Yerevan Physics Institute, Yerevan, Armenia}\\*[0pt]
A.M.~Sirunyan, A.~Tumasyan
\vskip\cmsinstskip
\textbf{Institut f\"{u}r Hochenergiephysik, Wien, Austria}\\*[0pt]
W.~Adam, F.~Ambrogi, E.~Asilar, T.~Bergauer, J.~Brandstetter, M.~Dragicevic, J.~Er\"{o}, A.~Escalante~Del~Valle, M.~Flechl, R.~Fr\"{u}hwirth\cmsAuthorMark{1}, V.M.~Ghete, J.~Hrubec, M.~Jeitler\cmsAuthorMark{1}, N.~Krammer, I.~Kr\"{a}tschmer, D.~Liko, T.~Madlener, I.~Mikulec, N.~Rad, H.~Rohringer, J.~Schieck\cmsAuthorMark{1}, R.~Sch\"{o}fbeck, M.~Spanring, D.~Spitzbart, W.~Waltenberger, J.~Wittmann, C.-E.~Wulz\cmsAuthorMark{1}, M.~Zarucki
% ...
\vskip\cmsinstskip
% ...
1:  Also at Vienna University of Technology, Vienna, Austria\\
% ...
\end{sloppypar}
\end{document}